\documentclass[aps,prl,twocolumn,superscriptaddress,showpacs]{revtex4-2}
\usepackage{amsmath,amssymb,wasysym,graphicx}
\usepackage[utf8]{inputenc}
\usepackage[T1]{fontenc}
\usepackage{xcolor}
\newcommand{\addYW}[1]{\textcolor{blue}{#1}}
\usepackage{textcomp}

\usepackage{bm}

\usepackage{color}
\definecolor{LinkColor}{rgb}{0.256,0.439,0.588}
 \usepackage{hyperref}
 \hypersetup{
    pdfauthor={good guys},
    pdftitle={good title},
    colorlinks=true,
    citecolor=LinkColor,
    linkcolor=LinkColor,
    urlcolor=LinkColor
 }

\usepackage{pifont}
\usepackage[normalem]{ulem}

\newcommand{\old}[1]{\textcolor{red}{\sout{#1}}}

\renewcommand{\old}[1]{}

\begin{document}

\title{The dynamical  exponent of
  a quantum critical itinerant ferromagnet: a Monte Carlo study}

\author{Yuzhi Liu}
\affiliation{Beijing National Laboratory for Condensed Matter Physics and Institute
	of Physics, Chinese Academy of Sciences, Beijing 100190, China}
\affiliation{School of Physical Sciences, University of Chinese Academy of Sciences, Beĳing 100190, China}

\author{Weilun Jiang}
\affiliation{Beijing National Laboratory for Condensed Matter Physics and Institute
	of Physics, Chinese Academy of Sciences, Beijing 100190, China}
\affiliation{School of Physical Sciences, University of Chinese Academy of Sciences, Beĳing 100190, China}

\author{Avraham Klein}
\affiliation{Department of Physics, Faculty of Natural Sciences, Ariel University, Ariel, Israel}

\author{Yuxuan Wang}
\affiliation{Department of Physics, University of Florida, Gainesville, FL 32601}

\author{Kai Sun}
\affiliation{Department of Physics, University of Michigan, Ann Arbor, MI 48109, USA}

\author{Andrey V. Chubukov}
\affiliation{School of Physics and Astronomy, University of Minnesota, Minneapolis, MN 55455, USA}

\author{Zi Yang Meng}
\email{zymeng@hku.hk}
\affiliation{Department of Physics and HKU-UCAS Joint Institute of Theoretical and Computational Physics, The University of Hong Kong, Pokfulam Road, Hong Kong SAR, China}
\affiliation{Beijing National Laboratory for Condensed Matter Physics and Institute of Physics, Chinese Academy of Sciences, Beijing 100190, China}

\date{\today}

\begin{abstract}
We consider the effect of the coupling between  2D quantum rotors near an XY  ferromagnetic quantum critical point and spins of itinerant fermions. We analyze how this coupling affects the dynamics of rotors and the self-energy of fermions.
  A common belief is that near a $q=0$ ferromagnetic transition, fermions induce an $\Omega/q$ Landau damping
  of rotors (i.e., the dynamical critical exponent is $z=3$) and Landau overdamped rotors give rise
  to non-Fermi liquid fermionic self-energy $\Sigma\propto \omega^{2/3}$.  This behavior has been confirmed in previous quantum Monte Carlo (QMC) studies.
  Here we show that for the XY case the behavior is different.  We report the results of
  large scale quantum Monte Carlo simulations,
  which show that at small frequencies $z=2$ and $\Sigma\propto \omega^{1/2}$. 
  We argue that the new behavior is associated with the fact that a fermionic  spin is by itself not a conserved quantity  due to spin-spin coupling to rotors, and a combination of self-energy and vertex corrections replaces  $1/q$ in the Landau damping by a constant.  We discuss the implication of these results to experiments.
 \end{abstract}

\maketitle

{\it Introduction}\,---\,
In the study of strongly correlated systems, quantum criticality in itinerant fermionic systems is of crucial importance, because it offers a pathway towards non-Fermi liquids and unconventional superconudctivity (See for example
Refs. ~\onlinecite{Lohneysen_rmp_2007, sachdev_quantum_2011, Lee_review_2018} and references therein).
In this study, we focus on ferromagnetic quantum critical points in itinerant fermion systems, where non-Fermi liquid (nFL) behaviors have been observed in the quantum critical region in a variety of materials, such as the Kondo lattice materials UGe$_2$~\cite{Huxley2003}, URhGe~\cite{Levy2005}, UCoGe~\cite{Stock2011}, YbNi$_4$P$_2$~\cite{Steppke2013} and more recently CeRh$_6$Ge$_4$~\cite{HQYuan2020,YiWu2021}, where in the latter a pressure-induced quantum critical point (QCP) with the characteristic power-law nFL specific heat and resistivity is reported. This experimental progress poses a series of theoretical questions on the origin and characterization of these nFL behaviors. In particular, it is of crucial importance to understand the fundamental principles that govern these QCPs and to identify the universal properties that are enforced by these principles.

On the theoretical side, extensive efforts have been devoted to this topic in the past few decades. Based on the Hertz-Millis-Moriya theory~\cite{Hertz1976,Millis1993,Moriya1978}, the dynamic critical exponent of an itinerant ferromagnetic QCP, or indeed any isotropic long-wavelength collective excitation with ordering vector $\mathbf{q}=0$, is $z=3$. The extension of the theory to study fermionic properties
\cite{Lee1989,Altshuler1994,Oganesyan2001,Abanov2003,Rech2006}, predicts that
fermions near such QCPs
are overdamped, with fermionic self-energy scaling as $\Sigma\propto\omega_n^{2/3}$, where $\omega_n$ represents the Matsubara frequency.
The fact that this power is less than $1$ implies that the system is an nFL
at low enough frequencies.
Within the one-loop framework,
these conclusions and scaling exponents are universal for all itinerant ferromagnetic QCPs.
When higher order contributions are taken into account, additional phenomena may appear,
e.g. first order behavior, spiral phases, and low-frequency scaling violations ~\cite{Kirkpatrick2003,Belitz2005,Rech2006,Maslov2009,Millis2006,Conduit2009,Metlitski2010a,Holder2015,Green2018}, as well as
superconductivity.
In particular,
if the order parameter (OP) is non-conserved, higher order processes
modify the damping of the bosons in the long-wavelength limit, and usually change the value of $z$ to $2$
~\cite{Mineev2013,Chubukov2014}.

With the recent development in quantum Monte Carlo (QMC) techniques\cite{HaoShi2016, LiuJunwei2017,XYXu2017self,Xu2017,WLJiang2019}, it has become possible to simulate such fermionic systems at large scale in the close vicinity of the quantum critical point~\cite{Berg2019,XYXu2019,WLJiang2021}.
Such simulations offer an unbiased and accurate numerical measurement to examine and to
test
these theoretical ideas.
In recent QMC studies on the itinerant $(2+1)$d ferromagnetic Ising quantum critical point, numerical results confirm the universal scaling relation predicted by the
$z=3$ theory~\cite{Xu2017,XYXu2020}.
The fermionic self-energy, properly extrapolated to $T=0$ ~\cite{XYXu2020,AKlein2020},
agrees with the
expected  non-Fermi liquid $\omega^{2/3}$ behavior at low energies.

In this Letter, we study
a ferromagnetic QCP in which the spin
OP is of XY type.
The key difference between an Ising-type and XY ferromagnetic QCP is that in the latter case
a spin of an itinerant fermion is not separately a conserved quantity in the presence of a
spin-spin interaction with a rotor, and the same is true for a rotor.

From large scale QMC simulations, we show that
instead of $z=3$ and $\Sigma\propto\omega_n^{2/3}$, the scaling exponent becomes $z=2$,
and fermion self-energy becomes
$\Sigma\propto\omega_n^{1/2}$.
The mechanism driving this change is the form of the boson damping $\propto |\Omega_n|/\Gamma(q,\Omega_n)$. We find that $\Gamma(q,\Omega_n)$ for conserved and non-conserved OPs is very different, as $\Gamma$ is constrained by a Ward identity in the conserved case only ~\cite{Mineev2013,Chubukov2009,Chubukov2014,Punk2016}. In our case we find $\Gamma(q,\Omega_n) \approx \Gamma_0$ is constant over a wide range of temperatures, frequencies and momenta, similar to 
that in
an \emph{antiferromagnet}. Such damping
arises from scattering processes beyond the one-loop order
and is generally associated with non-cancellation between self-energy and vertex corrections (including Aslamazov-Larkin-type terms)~\cite{Chubukov2014}.
Once
this
is introduced into the Hertz-Millis-Moriya framework, the bosonic dynamical exponent becomes $z=2$,
and the fermion self-energy gets modified to  $\Sigma\propto\omega_n^{1/2}$.

\begin{figure}[htp!]
\includegraphics[width=0.8\columnwidth]{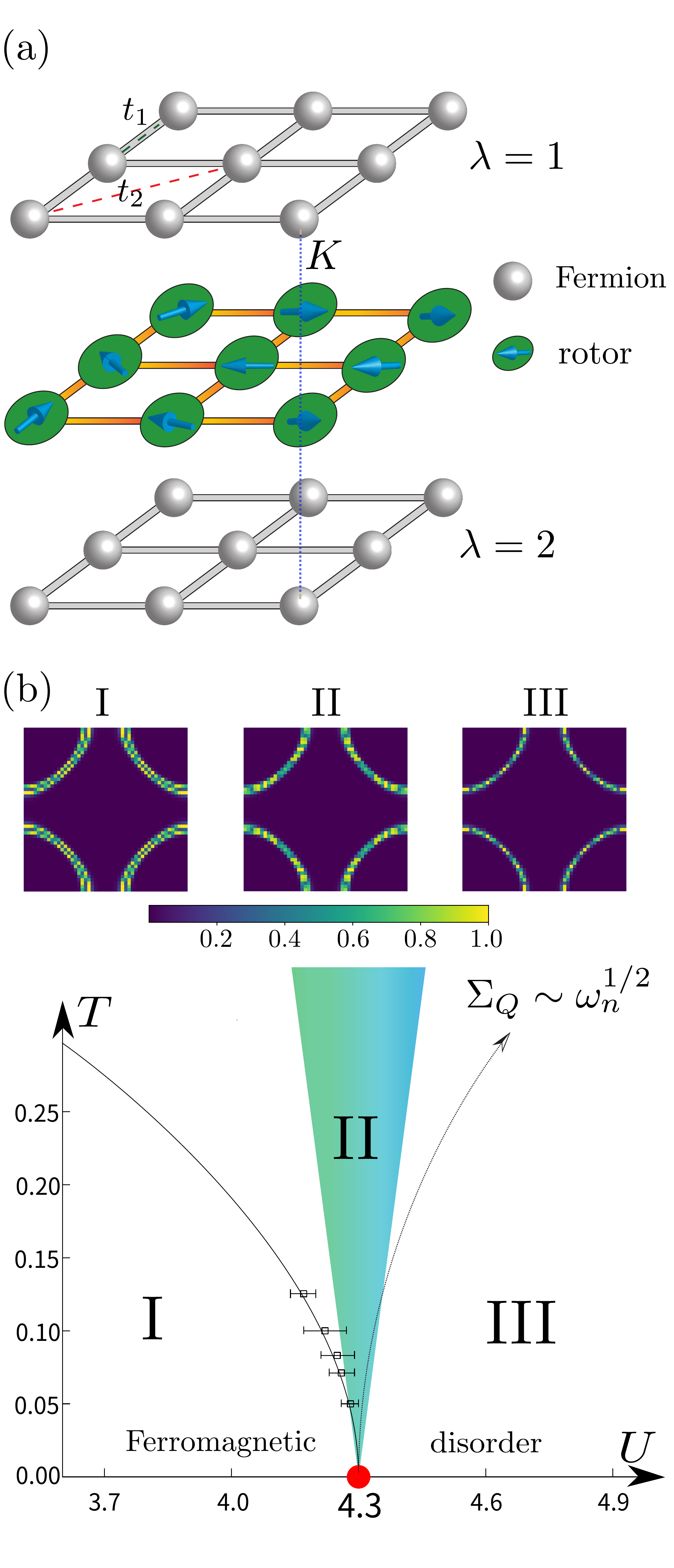}
\caption{Model and phase diagram. (a) The lattice model. The two identical layers of fermions ($\lambda=1,2$) with nearest-neighbor and next-nearest-neighbor hoppings $t_1=1$ and $t_2=0.2$ couple to the quantum rotor model in the middle layer with on-site coupling $K=1$.  As one tunes the rotors towards QCP, the entire system develops nFL behavior. (b) $T-U$ phase diagram. The QCP is located at $U_c=4.30(3)$, and when $U<U_c$ the system acquires ferromagnetic (quasi) long-range order
  below the $T_{BKT}$ boundary which extrapolates to $U_c$, as denoted by the black solid line, with the finite temperature transition points determined in SM~\cite{suppl}. Panels near the top part of (b) shows the Fermi surfaces, obtained from the dynamical Green's function $G(\bm{k},\tau=\beta/2)$ of $12\times 12$ size lattice with $\beta=1/T=24$, which correspond to the I(U=4.0), II(U=4.3) and III(U=4.5) regions of the phase diagram. In the  ferromagnetic phase, the Fermi surface splits. In the vicinity of the QCP, an nFL phase emerges due to strong quantum critical fluctuations, and the Fermi surface smears out. In the disordered region, the Fermi surface is close to that of the free system (see SM~\cite{suppl}).
}
\label{fig:fig1}
\end{figure}

{\it Model and Phase Diagram}\,---\,
We simulate lattice system composed of two identical fermion layers and one rotor layer as shown in Fig.~\ref{fig:fig1} (a), and the Hamiltonian is $\hat{H}=\hat{H}_f+\hat{H}_{qr}+\hat{H}_{int}.$ The fermion part of the Hamiltonian is $\hat{H}_{f}=-t_1\sum_{\left<i,j \right> \sigma,\lambda}\hat{c}_{i\sigma\lambda}^{\dagger}\hat{c}_{j\sigma\lambda}-t_2\sum_{\left<\left<i,j\right>\right>,\sigma,\lambda}\hat{c}_{i\sigma\lambda}^{\dagger}\hat{c}_{j\sigma\lambda} +h.c.,$ where $t_1=1$, $t_2=0.2$, $\langle\rangle$ ($\langle\langle\rangle\rangle$) denote (next) nearest neighbor, $\sigma=\uparrow$ or $\downarrow$ is the spin index, and $\lambda=1$ or $2$ labels the two fermion layers. For the rotor layer, we define a quantum rotor model (QRM) on the same square lattice with a Hamiltonian $\hat{H}_{qr}=\frac{U}{2}\sum_{i}\hat{L}^{2}_i-t_b\sum_{\langle i,j\rangle}\cos(\hat{\theta}_i-\hat{\theta}_j),$ where $\hat{L}_i$ and $\hat{\theta}_i$ are the angular momentum and polar angle of the rotor at site $i$ respectively. In the simulations, we set $t_b=1$ and use the ratio of $U/t_b$ to tune the system through the QCP. Without fermions, the phase diagram of the rotor model is well-known~\cite{WLJiang2019}. It contains two phases -- paramagnetic and ferromagnetic. At finite temperatures, the ferromagnetic phase shows quasi-long-range order and the thermal phase transition is Berezinskii-Kosterlitz-Thouless (BKT) type. At $T=0$, the ferromagnetic order becomes long-range and the quantum phase transition belongs to the $(2+1)$d XY universality class, which occurs at a QCP at $(U/t_b)_c=4.25(2)$~\cite{WLJiang2019}. The last term of the Hamiltonian $\hat{H}_{int}$  couples a  fermion spin ferromagnetically to
a quantum rotor at the same site:
\begin{align}
\hat{H}_{int} =-\frac{K}{2}\sum_{i}\hat{c}_{i}^{\dagger}\boldsymbol{\sigma}\hat{c}_{i}\cdot  \boldsymbol{\hat\theta}_i,
\end{align}
where $\bm{\sigma}$ represents fermion spin, and $ \boldsymbol{\hat\theta}_i =(\cos\theta_i,\sin\theta_i)$.
As we noted above,
this coupling term breaks the
spin symmetry for, separately, the fermions and the rotors,
replacing it with a rotation symmetry of the total spin (rotors + fermions). We set the coupling strength
to $K=1$.
We denote this model the \emph{XY-spin-fermion model}.

\old{For a larger coupling $K=4$, we found in a recent study~\cite{WLJiang2021} a pseudo-gap region and a superconducting dome
around the QCP, due to strong critical fluctuations. In this work,
we suppress superconductivity by utilizing a smaller value of $K$.
Because critical fluctuations induces an
attraction among fermions which scales as $K^2$, a smaller $K$, dramatically reduces this attraction and thus suppress superconducting transition temperature exponentially,
allowing us to examine the close vicinity of the QCP and its nontrivial critical scalings
in the normal state.}

In a recent study~\cite{WLJiang2021}, we used a similar model with $K=4$ to study the \emph{superconducting} properties of this model, and found a pseudo-gap region and a superconducting dome
around the QCP. While interesting on their own, these phenomena preempts the non-superconducting quantum-critical behavior of the system. In this work,
we suppress superconductivity by utilizing a smaller value of $K$,
driving the superconducting phase to unreachably low temperatures. This allows us to closely study the \emph{normal-state} critical properties in the vicinity of the QCP, and reveal a wealth of interesting features.

We plot the phase diagram of this model in Fig.~\ref{fig:fig1} (b). Similar to the QRM,the XY-spin-fermion system also exhibits two phases, paramagnetic and ferromagnetic, although the QCP now moves from $U_c=4.25(2)$ of the QRM to $U_c=4.30(3)$ here.
More importantly,  the presence of fermionic degrees of freedom has crucial impact on the quantum criticality, altering the dynamical exponent of the rotor propagator. In return, coupling to soft rotor
tends to make the fermions incoherent, with non-Fermi liquid self-energy.
\begin{figure}[htp!]
	\includegraphics[width=\columnwidth]{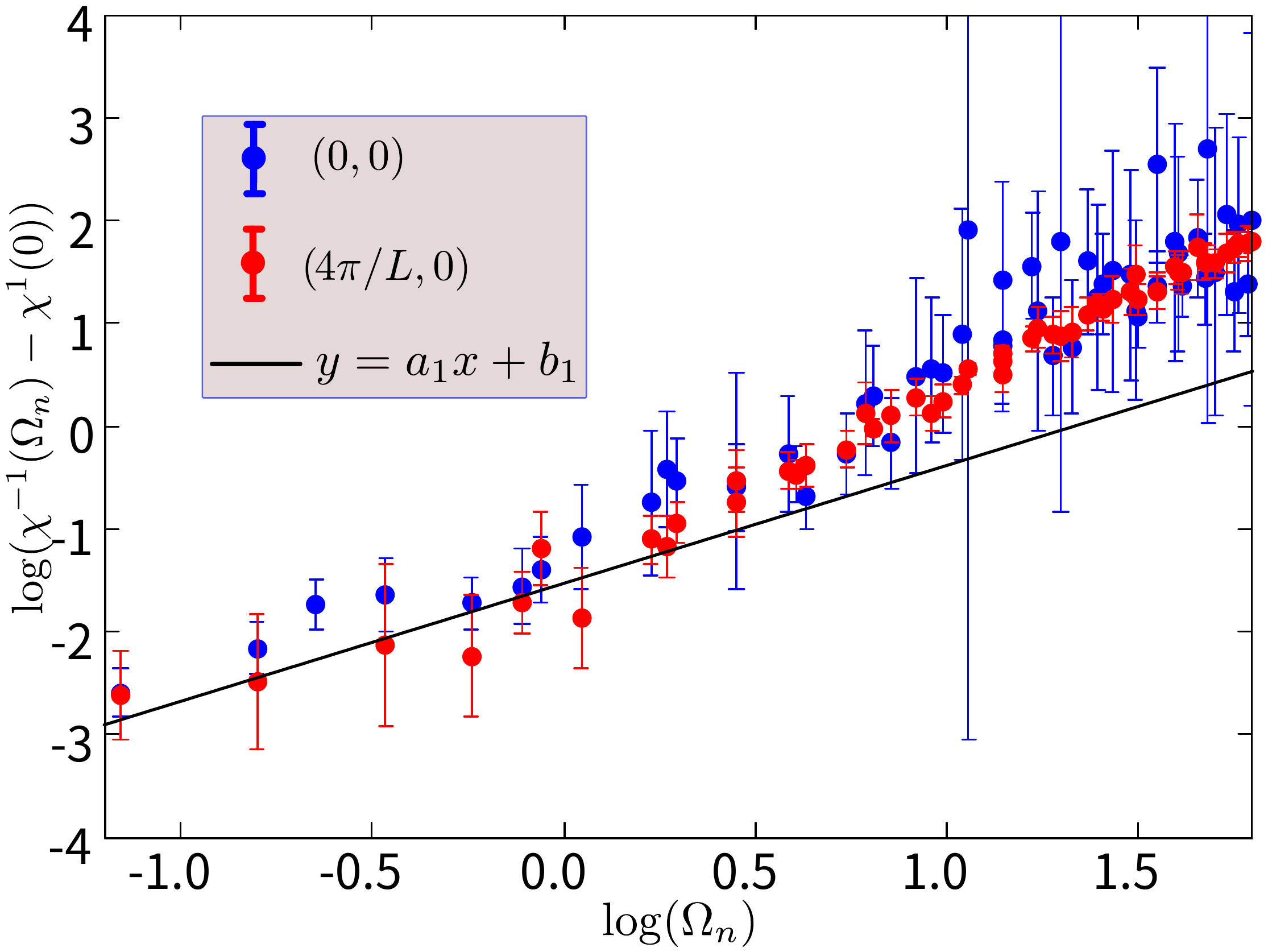}
	\caption{Inverse bosonic susceptibility versus frequency at the QCP with $\mathbf{q}=(0,0)$ and $(4 \pi/L,0)$. log-log plot for QMC data and fit the data in the range of $\text{log}(\Omega_n)<0$ by the black line. $\text{a}_1=1.14 \pm 0.2$, $\text{b}_1=-1.53 \pm 0.1$ are fitting parameters and $\text{a}_1$ is very close to $1$ which means the linear behavior at the small range $0<\Omega_n<1$.}
	\label{fig:fig2}
\end{figure}

{\it Results and Analysis}\,---\,
To study the scaling behavior of critical fluctuation, we measure the dynamic susceptibility of quantum rotors, $\chi\left(\mathbf{q},\Omega_n \right)=\frac{1}{L^2} \int d\tau \sum_{ij} e^{i \Omega_n \tau - i \mathbf{q} \mathbf{r}_{ij}}  \langle {\mathbf{\theta_i}}(\tau) {\mathbf{\theta_j}}(0)\rangle$ at $U_c=4.30(3)$ and low temperatures, where $\Omega_n = 2n\pi T$ is the bosonic Matsubara frequency. At small $\Omega_n$ and $q$, we find the momentum dependence of $\chi^{-1}$ to scale with $q^2$ (See supplemental material (SM)~\cite{suppl}) as expected. However, in the frequency dependence, we observe a completely different behavior from the prediction of Hertz-Millis-Moriya theory. 
For a system with a
  conserved
  OP, it is well-known that the Landau damping takes a singular form of $\frac{\Omega_n}{\sqrt{|\mathbf{q}|^2+\Omega_n^2}}$, and the $q=0$ susceptibility exhibits a discontinuity at zero frequency, i.e., $\lim_{\Omega_n\to 0}[\chi^{-1}(\mathbf{q}=0,\Omega_n) - \chi^{-1}(\mathbf{q}=0,\Omega_n=0)]$ is finite. This form is the base for $z=3$ dynamical critical exponent in the the Hertz-Millis-Moriya theory, which has been observed in QMC studies of the Ising QCP~\cite{Xu2017}.
  In contrast, our simulation exhibits no such singularity. Instead, as shown Fig.~\ref{fig:fig2} which contains two representative QMC results for $\mathbf{q}=(0,0)$ and ($4\pi/L$, 0), $\chi^{-1}(\mathbf{q},\Omega_n)$ are  smooth function of $\Omega_n$ without any discontinuity. This absence of singularity and discontinuity is our key observation, in direct contrast to the Hertz-Millis-Moriya theory as well as numerical results in the Ising-spin-fermion model~\cite{Xu2017}. Detailed data analysis reveals that within numerical uncertainty, $\chi^{-1}(0,\Omega_n)-\chi^{-1}(0,0)$ scales linearly with $\Omega_n$ at low frequency (the fit in Fig.~\ref{fig:fig2}), and thus the scaling behavior of the dynamic susceptibility indicates
  that $z=2$, analogous to
  an itinerant QCP in which an order
  breaks the  translational symmetry (e.g.~anti-ferromagnetic QCPs)~\cite{Liu2019, AKlein2020}.


We argue that this discrepancy is due to the non-conservation of the OP in the XY-spin-fermion model. At one-loop level,
the correction to a bosonic propagator comes from a polarization bubble of free fermions,
and the result
is the classic Landau damping
$\propto\frac{\Omega_n}{\sqrt{v_F^2|\mathbf{q}|^2+\Omega_n^2}}$.
For free fermions,
the Landau damping  arises whether or not the OP is conserved. Thus, at weak enough coupling, a discontinuity exists even for a nonconserved OP, as seen in e.g. simulations of nematic QCPs \cite{Schattner2016b}.
For a conserved OP,
this
form holds at all orders in perturbation theory due to
a Ward identity~\cite{Chubukov2005,Metlitski2010a,Chubukov2014,Klein2018}.
However, as we mentioned, in our XY-spin-fermion model, neither $\sigma^{x}$ nor $\sigma^{y}$ component of the fermion spin is conserved.
In this situation, vertex and self-energy corrections to fermion polarization due to spin-spin coupling to rotors
replace $1/\sqrt{v_F^2|\mathbf{q}|^2+\Omega_n^2}$ by a constant $\Gamma_0$,
giving rise to damping $\propto \Omega_n / \Gamma_0 + \mbox{corrections}$~\cite{Mineev2013,Chubukov2014}.
This changes the dynamical critical exponent to  $z=2$.

The change in $z$
has an important consequence for  the non-Fermi liquid fermion self-energy, which now must scale as
$\omega^{1/2}$, in analogy to
that for fermions
at the hotspots of an
anti-ferromagnetic QCP~\cite{Abanov2003,Liu2019}.  We verify this in in our QMC data.
Because
simulations are performed at finite temperature with discrete Matsubara frequencies, a thermal contribution (corresponding to processes with zero internal bosonic Matsubara frequency) needs to be deducted from the fermion self-energy, in order to expose the nFL behavior. Procedures for this deduction of thermal background have been developed in Refs.~\cite{AKlein2020,XYXu2020}, which we follow here (See SM~\cite{suppl} for details).
In the temperature range of our QMC simulations the fermionic self energy remains small, and the fermions remain in a Fermi liquid state, so that the  thermal contribution
to fermion self-energy
can be computed within Fermi liquid theory. It is
\begin{align}
\Sigma(k_F,\omega_n) =\Sigma_T(\omega_n)+\Sigma_Q(\omega_n)=\frac{\alpha}{\omega_n}+\Sigma_Q(\omega_n),
\end{align}
where $\omega_n$ is the Matsubara frequency and $\Sigma_T$ ($\Sigma_Q$) is the thermal (quantum) part of the self-energy.
The thermal part scales as $\Sigma_T\propto 1/\omega_n$, while the quantum part is the $T=0$ fermion self-energy,
 \begin{equation}
\Sigma_Q=\bar{g}\sigma(\omega_n)(\frac{\omega_n}{\omega_c})^{1/2}u(\frac{\omega_n}{\omega_c})
\label{eq:eq8}
\end{equation}
with $\sigma(\omega_n)$ being the sign function and
\begin{align}
	u(z) = \int_{0}^{\infty}\frac{dxdy}{4\pi^2}\frac{1}{x^2+y}\left(\frac{\sigma(y+1)}{\sqrt{1+(\frac{y+1}{x})^2 z^2}}-\frac{\sigma(y-1)}{\sqrt{1+(\frac{y-1}{x})^2 z^2}}\right)\nonumber
	\label{eq:eq9}
\end{align}
where the $\omega_c=\kappa \upsilon^2_f$ and $u(z)\rightarrow \frac{1}{2\pi}$ when $z \rightarrow 0$. The value of the coefficent $\bar{g}$ is given in the SM~\cite{suppl}.
At small $\omega$, the quantum part scales as $\Sigma_Q \propto \omega^{1/2}$.

\begin{figure}[tp!]
\includegraphics[width=\hsize]{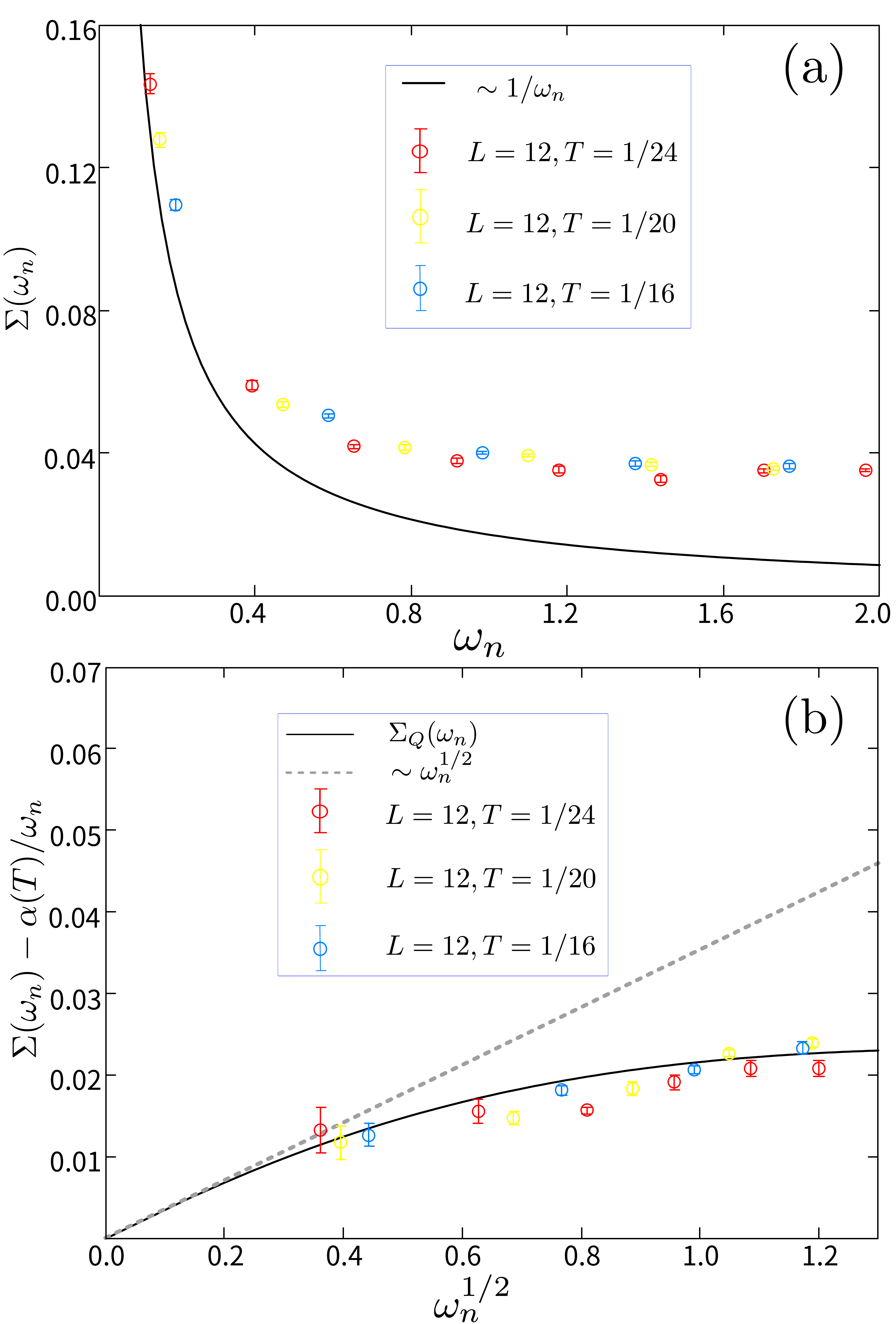}
\caption{Fermion self-energy. (a)
  $\Sigma(\mathbf{k}_F, \omega_n)$ from QMC at the QCP, here $\mathbf{k}_F$ is along the $(\pi,\pi)$ direction. The black line shows the thermal contribution, which scales as $\alpha/\omega_n,\alpha=0.01705$.
  (b) The quantum part of fermionic self-energy at QCP (after subtracting thermal contributions). The black line shows theory prediction of the zero-temperature  fermion self-energy [$\Sigma_{Q}(\omega_n)$], and the dashed line is its low-frequency asymptotic form of $~\omega^{1/2}$.}
\label{fig:fig3}
\end{figure}

We plot the fermion self-energy obtained from the QMC simulation at the QCP for the Fermi wavevector $\mathbf{k}_F$ along the diagonal direction (Fig.~\ref{fig:fig3}).
Because $\Sigma_T=\alpha /\omega_n$ and $\Sigma_Q\propto \sqrt{\omega_n}$, at low frequency, the self-energy is dominated by the thermal part. In Fig.~\ref{fig:fig3}.(a), indeed the low frequency data exhibits $1/\omega_n$ scaling (solid line), and the value of $\alpha$ can be obtained via numerical fitting.
In Fig.~\ref{fig:fig3} (b), we subtract the thermal part, utilizing this numerical fitted $\alpha$, and obtain
$\Sigma_Q$.
We also show the
theoretical prediction
for $\Sigma_Q$ [Eq.~\eqref{eq:eq8}], which agrees nicely with the QMC data.
We emphasize that the data analysis only utilizes one fitting parameter ($\alpha$), which is determined using only low frequency data points, while good agreement is obtained for a large frequency window. As mentioned early on, the quantum part is the fermion self-energy at $T=0$, and it scales as $\omega^{1/2}$ at low frequency, and this $\Sigma_Q\propto\omega^{1/2}$  asymptotic form is shown as the dashed line in Fig.~\ref{fig:fig3} (b).

{\it Discussion}\,---\,
We showed numerically that the ferromagnetic fluctuations have critical scaling $z=2$ and linear in frequency damping, giving rise to $\omega_n^{1/2}$ fermionic self energy.
The deviation from the expected $z=3$ scaling and $\omega_n^{2/3}$ self energy, is because Landau damping results from a delicate cancellation between scattering processes with different numbers of collective excitations, i.e. different loop order in a diagrammatic expansion \cite{Chubukov2005,Chubukov2009,Maslov2016}. Such cancellations occur only for a conserved OP, which in our system is the total spin, not
separately the rotor spin or the fermion spin.
As a result, the damping term in the rotor propagator is $
\Omega_n / \Gamma_0$.
For a purely fermionic system with a non-conserved OP (e.g., a nematic one),
the dominant contribution to $\Gamma_0$  comes from
thermally broadened fermions,
in which case $\Gamma_0 \approx 2\Sigma_T$.
Such behavior has been seen in previous QMC studies \cite{WLJiang2021,AKlein2020}.
In the
present case,  a finite $\Gamma_0$ likely arises from the non-cancellation between scattering processes involving different numbers of rotor propagators.
\old{It is also worthwhile to point out that for
  the model studied in Ref.~\cite{WLJiang2021}
  , because it contains the same type of itinerant fermions and the same symmetry-breaking pattern, in general the QCP
  should belong to the same universality class as the model studied here. However, because the critical fluctuations in that model induces a superconducting dome that covers the QCP, the universal scalings relation reported in this study cannot be observed, unless superconductivity fluctuations are suppressed by some mechanisms.}
We note that
  the model studied in Ref.~\cite{WLJiang2021} with $K=4$
 and the model studied here with $K=1$ should in principle belong to the same universality class. However, the huge difference in energy scales between the models means we cannot conclusively connect the two phase diagrams, and we leave such an investigation to a further systematic study.s


On the experimental side, whether the OP is conserved (Ising like)
or not (XY like), depends on the structure (e.g. strength and sign) of the spin-orbit (SO) coupling. Thus, our results indicate that SO couplings play a crucial role here and dictate the scaling exponents of such QCPs, as well as the associated non-Fermi liquids.
In materials with strong
XY anisotropy, we expect, near a QCP,
$z=2$ and $\Sigma\propto\omega^{1/2}$.

The values of critical exponent have direct impact
on the bosonic contribution to the specific heat. In 2D, for QCPs with conserved OPs, critical fluctuations generates a sub-linear specific heat $C_V\propto T^{2/3}$ (with $z=3$)\cite{Oganesyan2001}, which dominates over the linear $T$ contribution from the fermions. For non-conserved OPs, because $z=2$, this specific heat scales as $C_V\propto T$ (up to logarithmic corrections), same as the fermion contribution. Thus, experimentally, these two universality classes can be distinguished. For 3D systems, this specific heat anomaly is $C_V\propto T \ln(1/T)$ for QCPs with  conserved OPs~\cite{Millis1993} and $C_V\propto T^{3/2}$ if the OP is non-conserved. Notice that $C_V\propto T^{3/2}$ is a subleading correction to the fermion contributions of $C_V \propto T$, and thus it can be easily distinguished from the $C_V\propto T \ln(1/T)$ anomaly of the conserved case. In addition to specific heat, critical fluctuations and non-Fermi-liquid behavior generate other experimental signatures, such as transport, spectroscopy, X-ray/neutron scatterings, magnetic resonance, etc. In QMC simulations, these physical observables can all be measured, utilizing analytic continuations to convert imaginary time and Matsubara frequencies to real time and frequencies~\cite{WLJiang2021,Sandvik2016,CKZhou2021}. Such calculations will be performed in future studies, which can provide important guidance and insights for experimental studies in variety of quantum magnets, such as UGe$_2$~\cite{Huxley2003}, URhGe~\cite{Levy2005}, UCoGe~\cite{Stock2011} and YbNi$_4$P$_2$~\cite{Steppke2013} and CeRh$_6$Ge$_4$~\cite{HQYuan2020,YiWu2021}.

\acknowledgments
We thank R.M. Fernandes, M.H. Christensen, Y. Schattner, X. Wang and E. Berg for valuable discussions.
YZL, WLJ and ZYM acknowledge support from the RGC of Hong Kong SAR of China (Grant Nos. 17303019, 17301420 and AoE/P-701/20) and the Strategic Priority Research Program of the Chinese Academy of Sciences (Grant No. XDB33000000). We thank the Center for Quantum Simulation Sciences in the Institute of Physics, Chinese Academy of Sciences, the Computational Initiative at the Faculty of Science and the Information Technology Services at the University of Hong Kong and the Tianhe supercomputing platforms at the National Supercomputer Centers in Tianjin and Guangzhou for their technical support and generous allocation of CPU time. YW acknowledges support from NSF under award number DMR-2045871.
The work by AVC  was supported by  the Office of Basic Energy Sciences, U.S. Department of Energy, under award  DE-SC0014402. AK and  AVC acknowledge the hospitality of KITP at UCSB, where part of the
work has been conducted. The research at KITP is supported by the
National Science Foundation under Grant No. NSF PHY-1748958.
\bibliographystyle{apsrev4-1}
\bibliography{main}

\begin{thebibliography}{49}%
\makeatletter
\providecommand \@ifxundefined [1]{%
 \@ifx{#1\undefined}
}%
\providecommand \@ifnum [1]{%
 \ifnum #1\expandafter \@firstoftwo
 \else \expandafter \@secondoftwo
 \fi
}%
\providecommand \@ifx [1]{%
 \ifx #1\expandafter \@firstoftwo
 \else \expandafter \@secondoftwo
 \fi
}%
\providecommand \natexlab [1]{#1}%
\providecommand \enquote  [1]{``#1''}%
\providecommand \bibnamefont  [1]{#1}%
\providecommand \bibfnamefont [1]{#1}%
\providecommand \citenamefont [1]{#1}%
\providecommand \href@noop [0]{\@secondoftwo}%
\providecommand \href [0]{\begingroup \@sanitize@url \@href}%
\providecommand \@href[1]{\@@startlink{#1}\@@href}%
\providecommand \@@href[1]{\endgroup#1\@@endlink}%
\providecommand \@sanitize@url [0]{\catcode `\\12\catcode `\$12\catcode
  `\&12\catcode `\#12\catcode `\^12\catcode `\_12\catcode `\%12\relax}%
\providecommand \@@startlink[1]{}%
\providecommand \@@endlink[0]{}%
\providecommand \url  [0]{\begingroup\@sanitize@url \@url }%
\providecommand \@url [1]{\endgroup\@href {#1}{\urlprefix }}%
\providecommand \urlprefix  [0]{URL }%
\providecommand \Eprint [0]{\href }%
\providecommand \doibase [0]{http://dx.doi.org/}%
\providecommand \selectlanguage [0]{\@gobble}%
\providecommand \bibinfo  [0]{\@secondoftwo}%
\providecommand \bibfield  [0]{\@secondoftwo}%
\providecommand \translation [1]{[#1]}%
\providecommand \BibitemOpen [0]{}%
\providecommand \bibitemStop [0]{}%
\providecommand \bibitemNoStop [0]{.\EOS\space}%
\providecommand \EOS [0]{\spacefactor3000\relax}%
\providecommand \BibitemShut  [1]{\csname bibitem#1\endcsname}%
\let\auto@bib@innerbib\@empty
\bibitem [{\citenamefont {L\"ohneysen}\ \emph {et~al.}(2007)\citenamefont
  {L\"ohneysen}, \citenamefont {Rosch}, \citenamefont {Vojta},\ and\
  \citenamefont {W\"olfle}}]{Lohneysen_rmp_2007}%
  \BibitemOpen
  \bibfield  {author} {\bibinfo {author} {\bibfnamefont {H.~v.}\ \bibnamefont
  {L\"ohneysen}}, \bibinfo {author} {\bibfnamefont {A.}~\bibnamefont {Rosch}},
  \bibinfo {author} {\bibfnamefont {M.}~\bibnamefont {Vojta}}, \ and\ \bibinfo
  {author} {\bibfnamefont {P.}~\bibnamefont {W\"olfle}},\ }\href {\doibase
  10.1103/RevModPhys.79.1015} {\bibfield  {journal} {\bibinfo  {journal} {Rev.
  Mod. Phys.}\ }\textbf {\bibinfo {volume} {79}},\ \bibinfo {pages} {1015}
  (\bibinfo {year} {2007})}\BibitemShut {NoStop}%
\bibitem [{\citenamefont {Sachdev}(2011)}]{sachdev_quantum_2011}%
  \BibitemOpen
  \bibfield  {author} {\bibinfo {author} {\bibfnamefont {S.}~\bibnamefont
  {Sachdev}},\ }\href {\doibase 10.1017/CBO9780511973765} {\emph {\bibinfo
  {title} {Quantum {Phase} {Transitions}}}},\ \bibinfo {edition} {2nd}\ ed.\
  (\bibinfo  {publisher} {Cambridge University Press},\ \bibinfo {address}
  {Cambridge},\ \bibinfo {year} {2011})\BibitemShut {NoStop}%
\bibitem [{\citenamefont {Lee}(2018)}]{Lee_review_2018}%
  \BibitemOpen
  \bibfield  {author} {\bibinfo {author} {\bibfnamefont {S.-S.}\ \bibnamefont
  {Lee}},\ }\href {\doibase 10.1146/annurev-conmatphys-031016-025531}
  {\bibfield  {journal} {\bibinfo  {journal} {Annual Review of Condensed Matter
  Physics}\ }\textbf {\bibinfo {volume} {9}},\ \bibinfo {pages} {227} (\bibinfo
  {year} {2018})},\ \Eprint
  {http://arxiv.org/abs/https://doi.org/10.1146/annurev-conmatphys-031016-025531}
  {https://doi.org/10.1146/annurev-conmatphys-031016-025531} \BibitemShut
  {NoStop}%
\bibitem [{\citenamefont {Huxley}\ \emph {et~al.}(2003)\citenamefont {Huxley},
  \citenamefont {Raymond},\ and\ \citenamefont {Ressouche}}]{Huxley2003}%
  \BibitemOpen
  \bibfield  {author} {\bibinfo {author} {\bibfnamefont {A.~D.}\ \bibnamefont
  {Huxley}}, \bibinfo {author} {\bibfnamefont {S.}~\bibnamefont {Raymond}}, \
  and\ \bibinfo {author} {\bibfnamefont {E.}~\bibnamefont {Ressouche}},\ }\href
  {\doibase 10.1103/PhysRevLett.91.207201} {\bibfield  {journal} {\bibinfo
  {journal} {Phys. Rev. Lett.}\ }\textbf {\bibinfo {volume} {91}},\ \bibinfo
  {pages} {207201} (\bibinfo {year} {2003})}\BibitemShut {NoStop}%
\bibitem [{\citenamefont {L{\'e}vy}\ \emph {et~al.}(2005)\citenamefont
  {L{\'e}vy}, \citenamefont {Sheikin}, \citenamefont {Grenier},\ and\
  \citenamefont {Huxley}}]{Levy2005}%
  \BibitemOpen
  \bibfield  {author} {\bibinfo {author} {\bibfnamefont {F.}~\bibnamefont
  {L{\'e}vy}}, \bibinfo {author} {\bibfnamefont {I.}~\bibnamefont {Sheikin}},
  \bibinfo {author} {\bibfnamefont {B.}~\bibnamefont {Grenier}}, \ and\
  \bibinfo {author} {\bibfnamefont {A.~D.}\ \bibnamefont {Huxley}},\ }\href
  {\doibase 10.1126/science.1115498} {\bibfield  {journal} {\bibinfo  {journal}
  {Science}\ }\textbf {\bibinfo {volume} {309}},\ \bibinfo {pages} {1343}
  (\bibinfo {year} {2005})}\BibitemShut {NoStop}%
\bibitem [{\citenamefont {Stock}\ \emph {et~al.}(2011)\citenamefont {Stock},
  \citenamefont {Sokolov}, \citenamefont {Bourges}, \citenamefont {Tobash},
  \citenamefont {Gofryk}, \citenamefont {Ronning}, \citenamefont {Bauer},
  \citenamefont {Rule},\ and\ \citenamefont {Huxley}}]{Stock2011}%
  \BibitemOpen
  \bibfield  {author} {\bibinfo {author} {\bibfnamefont {C.}~\bibnamefont
  {Stock}}, \bibinfo {author} {\bibfnamefont {D.~A.}\ \bibnamefont {Sokolov}},
  \bibinfo {author} {\bibfnamefont {P.}~\bibnamefont {Bourges}}, \bibinfo
  {author} {\bibfnamefont {P.~H.}\ \bibnamefont {Tobash}}, \bibinfo {author}
  {\bibfnamefont {K.}~\bibnamefont {Gofryk}}, \bibinfo {author} {\bibfnamefont
  {F.}~\bibnamefont {Ronning}}, \bibinfo {author} {\bibfnamefont {E.~D.}\
  \bibnamefont {Bauer}}, \bibinfo {author} {\bibfnamefont {K.~C.}\ \bibnamefont
  {Rule}}, \ and\ \bibinfo {author} {\bibfnamefont {A.~D.}\ \bibnamefont
  {Huxley}},\ }\href {\doibase 10.1103/PhysRevLett.107.187202} {\bibfield
  {journal} {\bibinfo  {journal} {Phys. Rev. Lett.}\ }\textbf {\bibinfo
  {volume} {107}},\ \bibinfo {pages} {187202} (\bibinfo {year}
  {2011})}\BibitemShut {NoStop}%
\bibitem [{\citenamefont {Steppke}\ \emph {et~al.}(2013)\citenamefont
  {Steppke}, \citenamefont {K{\"u}chler}, \citenamefont {Lausberg},
  \citenamefont {Lengyel}, \citenamefont {Steinke}, \citenamefont {Borth},
  \citenamefont {L{\"u}hmann}, \citenamefont {Krellner}, \citenamefont
  {Nicklas}, \citenamefont {Geibel}, \citenamefont {Steglich},\ and\
  \citenamefont {Brando}}]{Steppke2013}%
  \BibitemOpen
  \bibfield  {author} {\bibinfo {author} {\bibfnamefont {A.}~\bibnamefont
  {Steppke}}, \bibinfo {author} {\bibfnamefont {R.}~\bibnamefont
  {K{\"u}chler}}, \bibinfo {author} {\bibfnamefont {S.}~\bibnamefont
  {Lausberg}}, \bibinfo {author} {\bibfnamefont {E.}~\bibnamefont {Lengyel}},
  \bibinfo {author} {\bibfnamefont {L.}~\bibnamefont {Steinke}}, \bibinfo
  {author} {\bibfnamefont {R.}~\bibnamefont {Borth}}, \bibinfo {author}
  {\bibfnamefont {T.}~\bibnamefont {L{\"u}hmann}}, \bibinfo {author}
  {\bibfnamefont {C.}~\bibnamefont {Krellner}}, \bibinfo {author}
  {\bibfnamefont {M.}~\bibnamefont {Nicklas}}, \bibinfo {author} {\bibfnamefont
  {C.}~\bibnamefont {Geibel}}, \bibinfo {author} {\bibfnamefont
  {F.}~\bibnamefont {Steglich}}, \ and\ \bibinfo {author} {\bibfnamefont
  {M.}~\bibnamefont {Brando}},\ }\href {\doibase 10.1126/science.1230583}
  {\bibfield  {journal} {\bibinfo  {journal} {Science}\ }\textbf {\bibinfo
  {volume} {339}},\ \bibinfo {pages} {933} (\bibinfo {year} {2013})},\ \Eprint
  {http://arxiv.org/abs/https://science.sciencemag.org/content/339/6122/933.full.pdf}
  {https://science.sciencemag.org/content/339/6122/933.full.pdf} \BibitemShut
  {NoStop}%
\bibitem [{\citenamefont {Shen}\ \emph {et~al.}(2020)\citenamefont {Shen},
  \citenamefont {Zhang}, \citenamefont {Komijani}, \citenamefont {Nicklas},
  \citenamefont {Borth}, \citenamefont {Wang}, \citenamefont {Chen},
  \citenamefont {Nie}, \citenamefont {Li}, \citenamefont {Lu}, \citenamefont
  {Lee}, \citenamefont {Smidman}, \citenamefont {Steglich}, \citenamefont
  {Coleman},\ and\ \citenamefont {Yuan}}]{HQYuan2020}%
  \BibitemOpen
  \bibfield  {author} {\bibinfo {author} {\bibfnamefont {B.}~\bibnamefont
  {Shen}}, \bibinfo {author} {\bibfnamefont {Y.}~\bibnamefont {Zhang}},
  \bibinfo {author} {\bibfnamefont {Y.}~\bibnamefont {Komijani}}, \bibinfo
  {author} {\bibfnamefont {M.}~\bibnamefont {Nicklas}}, \bibinfo {author}
  {\bibfnamefont {R.}~\bibnamefont {Borth}}, \bibinfo {author} {\bibfnamefont
  {A.}~\bibnamefont {Wang}}, \bibinfo {author} {\bibfnamefont {Y.}~\bibnamefont
  {Chen}}, \bibinfo {author} {\bibfnamefont {Z.}~\bibnamefont {Nie}}, \bibinfo
  {author} {\bibfnamefont {R.}~\bibnamefont {Li}}, \bibinfo {author}
  {\bibfnamefont {X.}~\bibnamefont {Lu}}, \bibinfo {author} {\bibfnamefont
  {H.}~\bibnamefont {Lee}}, \bibinfo {author} {\bibfnamefont {M.}~\bibnamefont
  {Smidman}}, \bibinfo {author} {\bibfnamefont {F.}~\bibnamefont {Steglich}},
  \bibinfo {author} {\bibfnamefont {P.}~\bibnamefont {Coleman}}, \ and\
  \bibinfo {author} {\bibfnamefont {H.}~\bibnamefont {Yuan}},\ }\href {\doibase
  10.1038/s41586-020-2052-z} {\bibfield  {journal} {\bibinfo  {journal}
  {Nature}\ }\textbf {\bibinfo {volume} {579}},\ \bibinfo {pages} {51 }
  (\bibinfo {year} {2020})}\BibitemShut {NoStop}%
\bibitem [{\citenamefont {Wu}\ \emph {et~al.}(2021)\citenamefont {Wu},
  \citenamefont {Zhang}, \citenamefont {Du}, \citenamefont {Shen},
  \citenamefont {Zheng}, \citenamefont {Fang}, \citenamefont {Smidman},
  \citenamefont {Cao}, \citenamefont {Steglich}, \citenamefont {Yuan},
  \citenamefont {Denlinger},\ and\ \citenamefont {Liu}}]{YiWu2021}%
  \BibitemOpen
  \bibfield  {author} {\bibinfo {author} {\bibfnamefont {Y.}~\bibnamefont
  {Wu}}, \bibinfo {author} {\bibfnamefont {Y.}~\bibnamefont {Zhang}}, \bibinfo
  {author} {\bibfnamefont {F.}~\bibnamefont {Du}}, \bibinfo {author}
  {\bibfnamefont {B.}~\bibnamefont {Shen}}, \bibinfo {author} {\bibfnamefont
  {H.}~\bibnamefont {Zheng}}, \bibinfo {author} {\bibfnamefont
  {Y.}~\bibnamefont {Fang}}, \bibinfo {author} {\bibfnamefont {M.}~\bibnamefont
  {Smidman}}, \bibinfo {author} {\bibfnamefont {C.}~\bibnamefont {Cao}},
  \bibinfo {author} {\bibfnamefont {F.}~\bibnamefont {Steglich}}, \bibinfo
  {author} {\bibfnamefont {H.}~\bibnamefont {Yuan}}, \bibinfo {author}
  {\bibfnamefont {J.~D.}\ \bibnamefont {Denlinger}}, \ and\ \bibinfo {author}
  {\bibfnamefont {Y.}~\bibnamefont {Liu}},\ }\href {\doibase
  10.1103/PhysRevLett.126.216406} {\bibfield  {journal} {\bibinfo  {journal}
  {Phys. Rev. Lett.}\ }\textbf {\bibinfo {volume} {126}},\ \bibinfo {pages}
  {216406} (\bibinfo {year} {2021})}\BibitemShut {NoStop}%
\bibitem [{\citenamefont {Hertz}(1976)}]{Hertz1976}%
  \BibitemOpen
  \bibfield  {author} {\bibinfo {author} {\bibfnamefont {J.~A.}\ \bibnamefont
  {Hertz}},\ }\href {\doibase 10.1103/PhysRevB.14.1165} {\bibfield  {journal}
  {\bibinfo  {journal} {Physical Review B}\ }\textbf {\bibinfo {volume} {14}},\
  \bibinfo {pages} {1165} (\bibinfo {year} {1976})}\BibitemShut {NoStop}%
\bibitem [{\citenamefont {Millis}(1993)}]{Millis1993}%
  \BibitemOpen
  \bibfield  {author} {\bibinfo {author} {\bibfnamefont {A.~J.}\ \bibnamefont
  {Millis}},\ }\href {\doibase 10.1103/PhysRevB.48.7183} {\bibfield  {journal}
  {\bibinfo  {journal} {Physical Review B}\ }\textbf {\bibinfo {volume} {48}},\
  \bibinfo {pages} {7183} (\bibinfo {year} {1993})}\BibitemShut {NoStop}%
\bibitem [{\citenamefont {Moriya}\ and\ \citenamefont
  {Takahashi}(1978)}]{Moriya1978}%
  \BibitemOpen
  \bibfield  {author} {\bibinfo {author} {\bibfnamefont {T.}~\bibnamefont
  {Moriya}}\ and\ \bibinfo {author} {\bibfnamefont {Y.}~\bibnamefont
  {Takahashi}},\ }\href@noop {} {\bibfield  {journal} {\bibinfo  {journal} {Le
  Journal de Physique Colloques}\ }\textbf {\bibinfo {volume} {39}} (\bibinfo
  {year} {1978})}\BibitemShut {NoStop}%
\bibitem [{\citenamefont {Lee}(1989)}]{Lee1989}%
  \BibitemOpen
  \bibfield  {author} {\bibinfo {author} {\bibfnamefont {P.~A.}\ \bibnamefont
  {Lee}},\ }\href {\doibase 10.1103/PhysRevLett.63.680} {\bibfield  {journal}
  {\bibinfo  {journal} {Phys. Rev. Lett.}\ }\textbf {\bibinfo {volume} {63}},\
  \bibinfo {pages} {680} (\bibinfo {year} {1989})}\BibitemShut {NoStop}%
\bibitem [{\citenamefont {Altshuler}\ \emph {et~al.}(1994)\citenamefont
  {Altshuler}, \citenamefont {Ioffe},\ and\ \citenamefont
  {Millis}}]{Altshuler1994}%
  \BibitemOpen
  \bibfield  {author} {\bibinfo {author} {\bibfnamefont {B.~L.}\ \bibnamefont
  {Altshuler}}, \bibinfo {author} {\bibfnamefont {L.~B.}\ \bibnamefont
  {Ioffe}}, \ and\ \bibinfo {author} {\bibfnamefont {A.~J.}\ \bibnamefont
  {Millis}},\ }\href {\doibase 10.1103/PhysRevB.50.14048} {\bibfield  {journal}
  {\bibinfo  {journal} {Phys. Rev. B}\ }\textbf {\bibinfo {volume} {50}},\
  \bibinfo {pages} {14048} (\bibinfo {year} {1994})}\BibitemShut {NoStop}%
\bibitem [{\citenamefont {Oganesyan}\ \emph {et~al.}(2001)\citenamefont
  {Oganesyan}, \citenamefont {Kivelson},\ and\ \citenamefont
  {Fradkin}}]{Oganesyan2001}%
  \BibitemOpen
  \bibfield  {author} {\bibinfo {author} {\bibfnamefont {V.}~\bibnamefont
  {Oganesyan}}, \bibinfo {author} {\bibfnamefont {S.~A.}\ \bibnamefont
  {Kivelson}}, \ and\ \bibinfo {author} {\bibfnamefont {E.}~\bibnamefont
  {Fradkin}},\ }\href {\doibase 10.1103/PhysRevB.64.195109} {\bibfield
  {journal} {\bibinfo  {journal} {Phys. Rev. B}\ }\textbf {\bibinfo {volume}
  {64}},\ \bibinfo {pages} {195109} (\bibinfo {year} {2001})}\BibitemShut
  {NoStop}%
\bibitem [{\citenamefont {Abanov}\ \emph {et~al.}(2003)\citenamefont {Abanov},
  \citenamefont {Chubukov},\ and\ \citenamefont {Schmalian}}]{Abanov2003}%
  \BibitemOpen
  \bibfield  {author} {\bibinfo {author} {\bibfnamefont {A.}~\bibnamefont
  {Abanov}}, \bibinfo {author} {\bibfnamefont {A.~V.}\ \bibnamefont
  {Chubukov}}, \ and\ \bibinfo {author} {\bibfnamefont {J.}~\bibnamefont
  {Schmalian}},\ }\href {\doibase 10.1080/0001873021000057123} {\bibfield
  {journal} {\bibinfo  {journal} {Advances in Physics}\ }\textbf {\bibinfo
  {volume} {52}},\ \bibinfo {pages} {119} (\bibinfo {year} {2003})}\BibitemShut
  {NoStop}%
\bibitem [{\citenamefont {Rech}\ \emph {et~al.}(2006)\citenamefont {Rech},
  \citenamefont {P\'epin},\ and\ \citenamefont {Chubukov}}]{Rech2006}%
  \BibitemOpen
  \bibfield  {author} {\bibinfo {author} {\bibfnamefont {J.}~\bibnamefont
  {Rech}}, \bibinfo {author} {\bibfnamefont {C.}~\bibnamefont {P\'epin}}, \
  and\ \bibinfo {author} {\bibfnamefont {A.~V.}\ \bibnamefont {Chubukov}},\
  }\href {\doibase 10.1103/PhysRevB.74.195126} {\bibfield  {journal} {\bibinfo
  {journal} {Phys. Rev. B}\ }\textbf {\bibinfo {volume} {74}},\ \bibinfo
  {pages} {195126} (\bibinfo {year} {2006})}\BibitemShut {NoStop}%
\bibitem [{\citenamefont {Kirkpatrick}\ and\ \citenamefont
  {Belitz}(2003)}]{Kirkpatrick2003}%
  \BibitemOpen
  \bibfield  {author} {\bibinfo {author} {\bibfnamefont {T.~R.}\ \bibnamefont
  {Kirkpatrick}}\ and\ \bibinfo {author} {\bibfnamefont {D.}~\bibnamefont
  {Belitz}},\ }\href {\doibase 10.1103/PhysRevB.67.024419} {\bibfield
  {journal} {\bibinfo  {journal} {Phys. Rev. B}\ }\textbf {\bibinfo {volume}
  {67}},\ \bibinfo {pages} {024419} (\bibinfo {year} {2003})}\BibitemShut
  {NoStop}%
\bibitem [{\citenamefont {Belitz}\ \emph {et~al.}(2005)\citenamefont {Belitz},
  \citenamefont {Kirkpatrick},\ and\ \citenamefont {Vojta}}]{Belitz2005}%
  \BibitemOpen
  \bibfield  {author} {\bibinfo {author} {\bibfnamefont {D.}~\bibnamefont
  {Belitz}}, \bibinfo {author} {\bibfnamefont {T.~R.}\ \bibnamefont
  {Kirkpatrick}}, \ and\ \bibinfo {author} {\bibfnamefont {T.}~\bibnamefont
  {Vojta}},\ }\href {\doibase 10.1103/RevModPhys.77.579} {\bibfield  {journal}
  {\bibinfo  {journal} {Rev. Mod. Phys.}\ }\textbf {\bibinfo {volume} {77}},\
  \bibinfo {pages} {579} (\bibinfo {year} {2005})}\BibitemShut {NoStop}%
\bibitem [{\citenamefont {Maslov}\ and\ \citenamefont
  {Chubukov}(2009)}]{Maslov2009}%
  \BibitemOpen
  \bibfield  {author} {\bibinfo {author} {\bibfnamefont {D.~L.}\ \bibnamefont
  {Maslov}}\ and\ \bibinfo {author} {\bibfnamefont {A.~V.}\ \bibnamefont
  {Chubukov}},\ }\href@noop {} {\bibfield  {journal} {\bibinfo  {journal}
  {Phys. Rev. B}\ }\textbf {\bibinfo {volume} {79}},\ \bibinfo {pages} {075112}
  (\bibinfo {year} {2009})}\BibitemShut {NoStop}%
\bibitem [{\citenamefont {Chubukov}\ \emph {et~al.}(2006)\citenamefont
  {Chubukov}, \citenamefont {Maslov},\ and\ \citenamefont
  {Millis}}]{Millis2006}%
  \BibitemOpen
  \bibfield  {author} {\bibinfo {author} {\bibfnamefont {A.~V.}\ \bibnamefont
  {Chubukov}}, \bibinfo {author} {\bibfnamefont {D.~L.}\ \bibnamefont
  {Maslov}}, \ and\ \bibinfo {author} {\bibfnamefont {A.~J.}\ \bibnamefont
  {Millis}},\ }\href {\doibase 10.1103/PhysRevB.73.045128} {\bibfield
  {journal} {\bibinfo  {journal} {Phys. Rev. B}\ }\textbf {\bibinfo {volume}
  {73}},\ \bibinfo {pages} {045128} (\bibinfo {year} {2006})}\BibitemShut
  {NoStop}%
\bibitem [{\citenamefont {Conduit}\ \emph {et~al.}(2009)\citenamefont
  {Conduit}, \citenamefont {Green},\ and\ \citenamefont
  {Simons}}]{Conduit2009}%
  \BibitemOpen
  \bibfield  {author} {\bibinfo {author} {\bibfnamefont {G.~J.}\ \bibnamefont
  {Conduit}}, \bibinfo {author} {\bibfnamefont {A.~G.}\ \bibnamefont {Green}},
  \ and\ \bibinfo {author} {\bibfnamefont {B.~D.}\ \bibnamefont {Simons}},\
  }\href {\doibase 10.1103/PhysRevLett.103.207201} {\bibfield  {journal}
  {\bibinfo  {journal} {Phys. Rev. Lett.}\ }\textbf {\bibinfo {volume} {103}},\
  \bibinfo {pages} {207201} (\bibinfo {year} {2009})}\BibitemShut {NoStop}%
\bibitem [{\citenamefont {Metlitski}\ and\ \citenamefont
  {Sachdev}(2010)}]{Metlitski2010a}%
  \BibitemOpen
  \bibfield  {author} {\bibinfo {author} {\bibfnamefont {M.~A.}\ \bibnamefont
  {Metlitski}}\ and\ \bibinfo {author} {\bibfnamefont {S.}~\bibnamefont
  {Sachdev}},\ }\href {\doibase 10.1103/PhysRevB.82.075127} {\bibfield
  {journal} {\bibinfo  {journal} {Phys. Rev. B}\ }\textbf {\bibinfo {volume}
  {82}},\ \bibinfo {pages} {075127} (\bibinfo {year} {2010})}\BibitemShut
  {NoStop}%
\bibitem [{\citenamefont {Holder}\ and\ \citenamefont
  {Metzner}(2015)}]{Holder2015}%
  \BibitemOpen
  \bibfield  {author} {\bibinfo {author} {\bibfnamefont {T.}~\bibnamefont
  {Holder}}\ and\ \bibinfo {author} {\bibfnamefont {W.}~\bibnamefont
  {Metzner}},\ }\href {\doibase 10.1103/PhysRevB.92.041112} {\bibfield
  {journal} {\bibinfo  {journal} {Phys. Rev. B}\ }\textbf {\bibinfo {volume}
  {92}},\ \bibinfo {pages} {041112} (\bibinfo {year} {2015})}\BibitemShut
  {NoStop}%
\bibitem [{\citenamefont {Green}\ \emph {et~al.}(2018)\citenamefont {Green},
  \citenamefont {Conduit},\ and\ \citenamefont {Kr{\"u}ger}}]{Green2018}%
  \BibitemOpen
  \bibfield  {author} {\bibinfo {author} {\bibfnamefont {A.~G.}\ \bibnamefont
  {Green}}, \bibinfo {author} {\bibfnamefont {G.}~\bibnamefont {Conduit}}, \
  and\ \bibinfo {author} {\bibfnamefont {F.}~\bibnamefont {Kr{\"u}ger}},\
  }\href@noop {} {\bibfield  {journal} {\bibinfo  {journal} {Annual Review of
  Condensed Matter Physics}\ }\textbf {\bibinfo {volume} {9}},\ \bibinfo
  {pages} {59} (\bibinfo {year} {2018})}\BibitemShut {NoStop}%
\bibitem [{\citenamefont {Mineev}(2013)}]{Mineev2013}%
  \BibitemOpen
  \bibfield  {author} {\bibinfo {author} {\bibfnamefont {V.~P.}\ \bibnamefont
  {Mineev}},\ }\href {\doibase 10.1103/PhysRevB.88.224408} {\bibfield
  {journal} {\bibinfo  {journal} {Phys. Rev. B}\ }\textbf {\bibinfo {volume}
  {88}},\ \bibinfo {pages} {224408} (\bibinfo {year} {2013})}\BibitemShut
  {NoStop}%
\bibitem [{\citenamefont {Chubukov}\ \emph {et~al.}(2014)\citenamefont
  {Chubukov}, \citenamefont {Betouras},\ and\ \citenamefont
  {Efremov}}]{Chubukov2014}%
  \BibitemOpen
  \bibfield  {author} {\bibinfo {author} {\bibfnamefont {A.~V.}\ \bibnamefont
  {Chubukov}}, \bibinfo {author} {\bibfnamefont {J.~J.}\ \bibnamefont
  {Betouras}}, \ and\ \bibinfo {author} {\bibfnamefont {D.~V.}\ \bibnamefont
  {Efremov}},\ }\href {\doibase 10.1103/PhysRevLett.112.037202} {\bibfield
  {journal} {\bibinfo  {journal} {Phys. Rev. Lett.}\ }\textbf {\bibinfo
  {volume} {112}},\ \bibinfo {pages} {037202} (\bibinfo {year}
  {2014})}\BibitemShut {NoStop}%
\bibitem [{\citenamefont {Shi}\ and\ \citenamefont {Zhang}(2016)}]{HaoShi2016}%
  \BibitemOpen
  \bibfield  {author} {\bibinfo {author} {\bibfnamefont {H.}~\bibnamefont
  {Shi}}\ and\ \bibinfo {author} {\bibfnamefont {S.}~\bibnamefont {Zhang}},\
  }\href {\doibase 10.1103/PhysRevE.93.033303} {\bibfield  {journal} {\bibinfo
  {journal} {Phys. Rev. E}\ }\textbf {\bibinfo {volume} {93}},\ \bibinfo
  {pages} {033303} (\bibinfo {year} {2016})}\BibitemShut {NoStop}%
\bibitem [{\citenamefont {Liu}\ \emph {et~al.}(2017)\citenamefont {Liu},
  \citenamefont {Shen}, \citenamefont {Qi}, \citenamefont {Meng},\ and\
  \citenamefont {Fu}}]{LiuJunwei2017}%
  \BibitemOpen
  \bibfield  {author} {\bibinfo {author} {\bibfnamefont {J.}~\bibnamefont
  {Liu}}, \bibinfo {author} {\bibfnamefont {H.}~\bibnamefont {Shen}}, \bibinfo
  {author} {\bibfnamefont {Y.}~\bibnamefont {Qi}}, \bibinfo {author}
  {\bibfnamefont {Z.~Y.}\ \bibnamefont {Meng}}, \ and\ \bibinfo {author}
  {\bibfnamefont {L.}~\bibnamefont {Fu}},\ }\href {\doibase
  10.1103/PhysRevB.95.241104} {\bibfield  {journal} {\bibinfo  {journal} {Phys.
  Rev. B}\ }\textbf {\bibinfo {volume} {95}},\ \bibinfo {pages} {241104}
  (\bibinfo {year} {2017})}\BibitemShut {NoStop}%
\bibitem [{\citenamefont {Xu}\ \emph {et~al.}(2017{\natexlab{a}})\citenamefont
  {Xu}, \citenamefont {Qi}, \citenamefont {Liu}, \citenamefont {Fu},\ and\
  \citenamefont {Meng}}]{XYXu2017self}%
  \BibitemOpen
  \bibfield  {author} {\bibinfo {author} {\bibfnamefont {X.~Y.}\ \bibnamefont
  {Xu}}, \bibinfo {author} {\bibfnamefont {Y.}~\bibnamefont {Qi}}, \bibinfo
  {author} {\bibfnamefont {J.}~\bibnamefont {Liu}}, \bibinfo {author}
  {\bibfnamefont {L.}~\bibnamefont {Fu}}, \ and\ \bibinfo {author}
  {\bibfnamefont {Z.~Y.}\ \bibnamefont {Meng}},\ }\href {\doibase
  10.1103/PhysRevB.96.041119} {\bibfield  {journal} {\bibinfo  {journal} {Phys.
  Rev. B}\ }\textbf {\bibinfo {volume} {96}},\ \bibinfo {pages} {041119}
  (\bibinfo {year} {2017}{\natexlab{a}})}\BibitemShut {NoStop}%
\bibitem [{\citenamefont {Xu}\ \emph {et~al.}(2017{\natexlab{b}})\citenamefont
  {Xu}, \citenamefont {Sun}, \citenamefont {Schattner}, \citenamefont {Berg},\
  and\ \citenamefont {Meng}}]{Xu2017}%
  \BibitemOpen
  \bibfield  {author} {\bibinfo {author} {\bibfnamefont {X.~Y.}\ \bibnamefont
  {Xu}}, \bibinfo {author} {\bibfnamefont {K.}~\bibnamefont {Sun}}, \bibinfo
  {author} {\bibfnamefont {Y.}~\bibnamefont {Schattner}}, \bibinfo {author}
  {\bibfnamefont {E.}~\bibnamefont {Berg}}, \ and\ \bibinfo {author}
  {\bibfnamefont {Z.~Y.}\ \bibnamefont {Meng}},\ }\href {\doibase
  10.1103/PhysRevX.7.031058} {\bibfield  {journal} {\bibinfo  {journal} {Phys.
  Rev. X}\ }\textbf {\bibinfo {volume} {7}},\ \bibinfo {pages} {031058}
  (\bibinfo {year} {2017}{\natexlab{b}})}\BibitemShut {NoStop}%
\bibitem [{\citenamefont {{Jiang}}\ \emph {et~al.}(2019)\citenamefont
  {{Jiang}}, \citenamefont {{Pan}}, \citenamefont {{Liu}},\ and\ \citenamefont
  {{Meng}}}]{WLJiang2019}%
  \BibitemOpen
  \bibfield  {author} {\bibinfo {author} {\bibfnamefont {W.}~\bibnamefont
  {{Jiang}}}, \bibinfo {author} {\bibfnamefont {G.}~\bibnamefont {{Pan}}},
  \bibinfo {author} {\bibfnamefont {Y.}~\bibnamefont {{Liu}}}, \ and\ \bibinfo
  {author} {\bibfnamefont {Z.~Y.}\ \bibnamefont {{Meng}}},\ }\href@noop {}
  {\bibfield  {journal} {\bibinfo  {journal} {arXiv e-prints}\ ,\ \bibinfo
  {eid} {arXiv:1912.08229}} (\bibinfo {year} {2019})},\ \Eprint
  {http://arxiv.org/abs/1912.08229} {arXiv:1912.08229 [cond-mat.str-el]}
  \BibitemShut {NoStop}%
\bibitem [{\citenamefont {Berg}\ \emph {et~al.}(2019)\citenamefont {Berg},
  \citenamefont {Lederer}, \citenamefont {Schattner},\ and\ \citenamefont
  {Trebst}}]{Berg2019}%
  \BibitemOpen
  \bibfield  {author} {\bibinfo {author} {\bibfnamefont {E.}~\bibnamefont
  {Berg}}, \bibinfo {author} {\bibfnamefont {S.}~\bibnamefont {Lederer}},
  \bibinfo {author} {\bibfnamefont {Y.}~\bibnamefont {Schattner}}, \ and\
  \bibinfo {author} {\bibfnamefont {S.}~\bibnamefont {Trebst}},\ }\href@noop {}
  {\bibfield  {journal} {\bibinfo  {journal} {Annual Review of Condensed Matter
  Physics}\ }\textbf {\bibinfo {volume} {10}},\ \bibinfo {pages} {63} (\bibinfo
  {year} {2019})}\BibitemShut {NoStop}%
\bibitem [{\citenamefont {Xu}\ \emph {et~al.}(2019)\citenamefont {Xu},
  \citenamefont {Liu}, \citenamefont {Pan}, \citenamefont {Qi}, \citenamefont
  {Sun},\ and\ \citenamefont {Meng}}]{XYXu2019}%
  \BibitemOpen
  \bibfield  {author} {\bibinfo {author} {\bibfnamefont {X.~Y.}\ \bibnamefont
  {Xu}}, \bibinfo {author} {\bibfnamefont {Z.~H.}\ \bibnamefont {Liu}},
  \bibinfo {author} {\bibfnamefont {G.}~\bibnamefont {Pan}}, \bibinfo {author}
  {\bibfnamefont {Y.}~\bibnamefont {Qi}}, \bibinfo {author} {\bibfnamefont
  {K.}~\bibnamefont {Sun}}, \ and\ \bibinfo {author} {\bibfnamefont {Z.~Y.}\
  \bibnamefont {Meng}},\ }\href {\doibase 10.1088/1361-648x/ab3295} {\bibfield
  {journal} {\bibinfo  {journal} {Journal of Physics: Condensed Matter}\
  }\textbf {\bibinfo {volume} {31}},\ \bibinfo {pages} {463001} (\bibinfo
  {year} {2019})}\BibitemShut {NoStop}%
\bibitem [{\citenamefont {{Jiang}}\ \emph {et~al.}(2021)\citenamefont
  {{Jiang}}, \citenamefont {{Liu}}, \citenamefont {{Klein}}, \citenamefont
  {{Wang}}, \citenamefont {{Sun}}, \citenamefont {{Chubukov}},\ and\
  \citenamefont {{Meng}}}]{WLJiang2021}%
  \BibitemOpen
  \bibfield  {author} {\bibinfo {author} {\bibfnamefont {W.}~\bibnamefont
  {{Jiang}}}, \bibinfo {author} {\bibfnamefont {Y.}~\bibnamefont {{Liu}}},
  \bibinfo {author} {\bibfnamefont {A.}~\bibnamefont {{Klein}}}, \bibinfo
  {author} {\bibfnamefont {Y.}~\bibnamefont {{Wang}}}, \bibinfo {author}
  {\bibfnamefont {K.}~\bibnamefont {{Sun}}}, \bibinfo {author} {\bibfnamefont
  {A.~V.}\ \bibnamefont {{Chubukov}}}, \ and\ \bibinfo {author} {\bibfnamefont
  {Z.~Y.}\ \bibnamefont {{Meng}}},\ }\href@noop {} {\bibfield  {journal}
  {\bibinfo  {journal} {arXiv e-prints}\ ,\ \bibinfo {eid} {arXiv:2105.03639}}
  (\bibinfo {year} {2021})},\ \Eprint {http://arxiv.org/abs/2105.03639}
  {arXiv:2105.03639 [cond-mat.str-el]} \BibitemShut {NoStop}%
\bibitem [{\citenamefont {Xu}\ \emph {et~al.}(2020)\citenamefont {Xu},
  \citenamefont {Klein}, \citenamefont {Sun}, \citenamefont {Chubukov},\ and\
  \citenamefont {Meng}}]{XYXu2020}%
  \BibitemOpen
  \bibfield  {author} {\bibinfo {author} {\bibfnamefont {X.~Y.}\ \bibnamefont
  {Xu}}, \bibinfo {author} {\bibfnamefont {A.}~\bibnamefont {Klein}}, \bibinfo
  {author} {\bibfnamefont {K.}~\bibnamefont {Sun}}, \bibinfo {author}
  {\bibfnamefont {A.~V.}\ \bibnamefont {Chubukov}}, \ and\ \bibinfo {author}
  {\bibfnamefont {Z.~Y.}\ \bibnamefont {Meng}},\ }\href {\doibase
  10.1038/s41535-020-00266-6} {\bibfield  {journal} {\bibinfo  {journal} {npj
  Quantum Materials}\ }\textbf {\bibinfo {volume} {5}},\ \bibinfo {pages} {65}
  (\bibinfo {year} {2020})}\BibitemShut {NoStop}%
\bibitem [{\citenamefont {Klein}\ \emph {et~al.}(2020)\citenamefont {Klein},
  \citenamefont {Chubukov}, \citenamefont {Schattner},\ and\ \citenamefont
  {Berg}}]{AKlein2020}%
  \BibitemOpen
  \bibfield  {author} {\bibinfo {author} {\bibfnamefont {A.}~\bibnamefont
  {Klein}}, \bibinfo {author} {\bibfnamefont {A.~V.}\ \bibnamefont {Chubukov}},
  \bibinfo {author} {\bibfnamefont {Y.}~\bibnamefont {Schattner}}, \ and\
  \bibinfo {author} {\bibfnamefont {E.}~\bibnamefont {Berg}},\ }\href {\doibase
  10.1103/PhysRevX.10.031053} {\bibfield  {journal} {\bibinfo  {journal} {Phys.
  Rev. X}\ }\textbf {\bibinfo {volume} {10}},\ \bibinfo {pages} {031053}
  (\bibinfo {year} {2020})}\BibitemShut {NoStop}%
\bibitem [{\citenamefont {Chubukov}\ and\ \citenamefont
  {Maslov}(2009)}]{Chubukov2009}%
  \BibitemOpen
  \bibfield  {author} {\bibinfo {author} {\bibfnamefont {A.~V.}\ \bibnamefont
  {Chubukov}}\ and\ \bibinfo {author} {\bibfnamefont {D.~L.}\ \bibnamefont
  {Maslov}},\ }\href {\doibase 10.1103/PhysRevLett.103.216401} {\bibfield
  {journal} {\bibinfo  {journal} {Phys. Rev. Lett.}\ }\textbf {\bibinfo
  {volume} {103}},\ \bibinfo {pages} {216401} (\bibinfo {year}
  {2009})}\BibitemShut {NoStop}%
\bibitem [{\citenamefont {Punk}(2016)}]{Punk2016}%
  \BibitemOpen
  \bibfield  {author} {\bibinfo {author} {\bibfnamefont {M.}~\bibnamefont
  {Punk}},\ }\href {\doibase 10.1103/PhysRevB.94.195113} {\bibfield  {journal}
  {\bibinfo  {journal} {Phys. Rev. B}\ }\textbf {\bibinfo {volume} {94}},\
  \bibinfo {pages} {195113} (\bibinfo {year} {2016})}\BibitemShut {NoStop}%
\bibitem [{sup()}]{suppl}%
  \BibitemOpen
  \href@noop {} {\bibinfo  {journal} {See Supplemental Materal for QMC
  implementation of the model and the fitting of the QMC data are given, which
  includes Refs.\cite{WLJiang2019,BSS1981,Xu2017,WLJiang2021,Assaad2002,
  Chubukov2014,XYXu2020, AKlein2020}}\ }\BibitemShut {NoStop}%
\bibitem [{\citenamefont {Liu}\ \emph {et~al.}(2019)\citenamefont {Liu},
  \citenamefont {Pan}, \citenamefont {Xu}, \citenamefont {Sun},\ and\
  \citenamefont {Meng}}]{Liu2019}%
  \BibitemOpen
\bibfield  {journal} {  }\bibfield  {author} {\bibinfo {author} {\bibfnamefont
  {Z.~H.}\ \bibnamefont {Liu}}, \bibinfo {author} {\bibfnamefont
  {G.}~\bibnamefont {Pan}}, \bibinfo {author} {\bibfnamefont {X.~Y.}\
  \bibnamefont {Xu}}, \bibinfo {author} {\bibfnamefont {K.}~\bibnamefont
  {Sun}}, \ and\ \bibinfo {author} {\bibfnamefont {Z.~Y.}\ \bibnamefont
  {Meng}},\ }\href {\doibase 10.1073/pnas.1901751116} {\bibfield  {journal}
  {\bibinfo  {journal} {Proc Natl Acad Sci U S A}\ }\textbf {\bibinfo {volume}
  {116}},\ \bibinfo {pages} {16760} (\bibinfo {year} {2019})}\BibitemShut
  {NoStop}%
\bibitem [{\citenamefont {Schattner}\ \emph {et~al.}(2016)\citenamefont
  {Schattner}, \citenamefont {Lederer}, \citenamefont {Kivelson},\ and\
  \citenamefont {Berg}}]{Schattner2016b}%
  \BibitemOpen
  \bibfield  {author} {\bibinfo {author} {\bibfnamefont {Y.}~\bibnamefont
  {Schattner}}, \bibinfo {author} {\bibfnamefont {S.}~\bibnamefont {Lederer}},
  \bibinfo {author} {\bibfnamefont {S.~A.}\ \bibnamefont {Kivelson}}, \ and\
  \bibinfo {author} {\bibfnamefont {E.}~\bibnamefont {Berg}},\ }\href {\doibase
  10.1103/PhysRevX.6.031028} {\bibfield  {journal} {\bibinfo  {journal}
  {Physical Review X}\ }\textbf {\bibinfo {volume} {6}} (\bibinfo {year}
  {2016}),\ 10.1103/PhysRevX.6.031028}\BibitemShut {NoStop}%
\bibitem [{\citenamefont {Chubukov}(2005)}]{Chubukov2005}%
  \BibitemOpen
  \bibfield  {author} {\bibinfo {author} {\bibfnamefont {A.~V.}\ \bibnamefont
  {Chubukov}},\ }\href {\doibase 10.1103/PhysRevB.72.085113} {\bibfield
  {journal} {\bibinfo  {journal} {Phys. Rev. B}\ }\textbf {\bibinfo {volume}
  {72}},\ \bibinfo {pages} {085113} (\bibinfo {year} {2005})}\BibitemShut
  {NoStop}%
\bibitem [{\citenamefont {Klein}\ \emph {et~al.}(2018)\citenamefont {Klein},
  \citenamefont {Lederer}, \citenamefont {Chowdhury}, \citenamefont {Berg},\
  and\ \citenamefont {Chubukov}}]{Klein2018}%
  \BibitemOpen
  \bibfield  {author} {\bibinfo {author} {\bibfnamefont {A.}~\bibnamefont
  {Klein}}, \bibinfo {author} {\bibfnamefont {S.}~\bibnamefont {Lederer}},
  \bibinfo {author} {\bibfnamefont {D.}~\bibnamefont {Chowdhury}}, \bibinfo
  {author} {\bibfnamefont {E.}~\bibnamefont {Berg}}, \ and\ \bibinfo {author}
  {\bibfnamefont {A.}~\bibnamefont {Chubukov}},\ }\href {\doibase
  10.1103/PhysRevB.97.155115} {\bibfield  {journal} {\bibinfo  {journal} {Phys.
  Rev. B}\ }\textbf {\bibinfo {volume} {97}},\ \bibinfo {pages} {155115}
  (\bibinfo {year} {2018})}\BibitemShut {NoStop}%
\bibitem [{\citenamefont {Maslov}\ and\ \citenamefont
  {Chubukov}(2016)}]{Maslov2016}%
  \BibitemOpen
  \bibfield  {author} {\bibinfo {author} {\bibfnamefont {D.~L.}\ \bibnamefont
  {Maslov}}\ and\ \bibinfo {author} {\bibfnamefont {A.~V.}\ \bibnamefont
  {Chubukov}},\ }\href {\doibase 10.1088/1361-6633/80/2/026503} {\bibfield
  {journal} {\bibinfo  {journal} {Reports on Progress in Physics}\ }\textbf
  {\bibinfo {volume} {80}},\ \bibinfo {pages} {026503} (\bibinfo {year}
  {2016})}\BibitemShut {NoStop}%
\bibitem [{\citenamefont {Sandvik}(2016)}]{Sandvik2016}%
  \BibitemOpen
  \bibfield  {author} {\bibinfo {author} {\bibfnamefont {A.~W.}\ \bibnamefont
  {Sandvik}},\ }\href {\doibase 10.1103/PhysRevE.94.063308} {\bibfield
  {journal} {\bibinfo  {journal} {Phys. Rev. E}\ }\textbf {\bibinfo {volume}
  {94}},\ \bibinfo {pages} {063308} (\bibinfo {year} {2016})}\BibitemShut
  {NoStop}%
\bibitem [{\citenamefont {Zhou}\ \emph {et~al.}(2021)\citenamefont {Zhou},
  \citenamefont {Yan}, \citenamefont {Wu}, \citenamefont {Sun}, \citenamefont
  {Starykh},\ and\ \citenamefont {Meng}}]{CKZhou2021}%
  \BibitemOpen
  \bibfield  {author} {\bibinfo {author} {\bibfnamefont {C.}~\bibnamefont
  {Zhou}}, \bibinfo {author} {\bibfnamefont {Z.}~\bibnamefont {Yan}}, \bibinfo
  {author} {\bibfnamefont {H.-Q.}\ \bibnamefont {Wu}}, \bibinfo {author}
  {\bibfnamefont {K.}~\bibnamefont {Sun}}, \bibinfo {author} {\bibfnamefont
  {O.~A.}\ \bibnamefont {Starykh}}, \ and\ \bibinfo {author} {\bibfnamefont
  {Z.~Y.}\ \bibnamefont {Meng}},\ }\href {\doibase
  10.1103/PhysRevLett.126.227201} {\bibfield  {journal} {\bibinfo  {journal}
  {Phys. Rev. Lett.}\ }\textbf {\bibinfo {volume} {126}},\ \bibinfo {pages}
  {227201} (\bibinfo {year} {2021})}\BibitemShut {NoStop}%
\bibitem [{\citenamefont {Blankenbecler}\ \emph {et~al.}(1981)\citenamefont
  {Blankenbecler}, \citenamefont {Scalapino},\ and\ \citenamefont
  {Sugar}}]{BSS1981}%
  \BibitemOpen
  \bibfield  {author} {\bibinfo {author} {\bibfnamefont {R.}~\bibnamefont
  {Blankenbecler}}, \bibinfo {author} {\bibfnamefont {D.~J.}\ \bibnamefont
  {Scalapino}}, \ and\ \bibinfo {author} {\bibfnamefont {R.~L.}\ \bibnamefont
  {Sugar}},\ }\href {\doibase 10.1103/PhysRevD.24.2278} {\bibfield  {journal}
  {\bibinfo  {journal} {Phys. Rev. D}\ }\textbf {\bibinfo {volume} {24}},\
  \bibinfo {pages} {2278} (\bibinfo {year} {1981})}\BibitemShut {NoStop}%
\bibitem [{\citenamefont {Assaad}(2002)}]{Assaad2002}%
  \BibitemOpen
  \bibfield  {author} {\bibinfo {author} {\bibfnamefont {F.~F.}\ \bibnamefont
  {Assaad}},\ }\href {\doibase 10.1103/PhysRevB.65.115104} {\bibfield
  {journal} {\bibinfo  {journal} {Phys. Rev. B}\ }\textbf {\bibinfo {volume}
  {65}},\ \bibinfo {pages} {115104} (\bibinfo {year} {2002})}\BibitemShut
  {NoStop}%
\end{thebibliography}%


%
\clearpage
\onecolumngrid
\begin{center}
\textbf{Supplemental Material for "The dynamical  exponent of a quantum critical itinerant ferromagnet: a Monte Carlo study"}
\end{center}
\setcounter{equation}{0}
\setcounter{figure}{0}
\setcounter{table}{0}
\setcounter{page}{1}
\makeatletter
\renewcommand{\thetable}{S\arabic{table}}
\renewcommand{\theequation}{S\arabic{equation}}
\renewcommand{\thefigure}{S\arabic{figure}}
\setcounter{secnumdepth}{3}

\section{Quantum Rotor Model}
We start the Monte Carlo simulation for quantum rotor model (QRM). The Hamitonian is written as  
\begin{equation}
H_{\text{qr}}=\hat{T} + \hat{U} =\frac{U}{2} \sum_{i} (-i\frac{\partial}{\partial \hat{\theta}})^{2} - t_b\sum_{\left\langle i,j \right\rangle}\cos(\hat{\theta}_i-\hat{\theta}_j). 
 \label{eq:eqs1}
\end{equation}
The Hamitonian is shown in the $\hat{\theta}$ representation which the $\theta$ variable is located in each site, ranging between $[0,2\pi)$ and the partition function is
\begin{equation}
Z = \text{Tr} \left[ \exp (-\beta\big(-\frac{U}{2}\sum_i\frac{\partial^2}{\partial \hat \theta^{2}_{i}}-t_b\sum_{\langle i,j \rangle}\cos(\hat \theta_i - \hat \theta_j))\big) \right]
\end{equation} \label{eq:eqs2}

Under the Trotter decomposition. The $\beta$ is divided into $M$ slices with step $\Delta\tau = \beta/M$ and insert the complete sets of the $\left\{\theta_i\right \}$ on the each time slice. We have
\begin{equation}
   Z = \int \mathcal{D}\theta \prod_{l=0}^{M-1} \langle\{\theta(l+1)\}| e^{-\Delta\tau \hat T} e^{-\Delta\tau \hat V}|\{\theta(l)\}\rangle
\end{equation} \label{eq:eqs3}
The periodic boundary condition is $\{\theta(M)\} = \{\theta(0)\}$. The potential energy part can be straightforwardly treated since it is diagonal in the basis of  $\theta_i(l)$. For the kinetic energy part, we insert a complete set of the angular momentum eigenstates $|J_i(l)\rangle$ at site $i$ and time slice $l$:
\begin{equation}
   T(l) = \sum_{\{J\}}\prod_{i} e^{-\frac{\Delta\tau U}{2}[J_i(l)]^2} \langle \theta_i(l+1)|J_i(l)\rangle \langle J_i(l)|\theta_i(l)\rangle,
\end{equation} \label{eq:eqs4}
The term $\langle \theta_i(l) | J_i(l) \rangle$ equals to a complex value $e^{i J_i(l) \theta_i(l)}$. With the Poisson summation formula, we write
\begin{eqnarray}
T(l) &=& \prod_{i} \sum_{J}e^{-\frac{\Delta\tau U}{2} J^2}e^{iJ(\theta_i(l)-\theta_i(l+1))} \nonumber\\
&=& \prod_{i} \sum_{m=-\infty}^{\infty}\int_{-\infty}^{\infty}dJ e^{2\pi i J m}e^{-\frac{\Delta\tau U}{2}J^2}e^{iJ(\theta_i(l)-\theta_i(l+1))} \nonumber\\
&=& \prod_{i} \sum_{m=-\infty}^{\infty}\sqrt{\frac{2\pi}{\Delta\tau U}} e^{-\frac{1}{2\Delta\tau U}(\theta_i(l)-\theta_i(l+1) - 2\pi m)^2}.
\label{eq:eqs5}
\end{eqnarray} 
with Villian approximation 
\begin{equation}
T(l) \approx \prod_{i} e^{\frac{1}{\Delta \tau U}\cos(\theta_i(l)-\theta_i(l+1))}
\label{eq:eqs6}
\end{equation}
Using this result, we map the $2d$-QRM to $2d+1$ anisotropic XY model~\cite{WLJiang2019} with the partition function

\begin{eqnarray}
Z&=&\text{Tr} \left\{\int \mathcal{D}\theta \prod_{l=0}^{M-1} \langle\{\theta(l+1)\} | e^{-\Delta \tau \hat H_{\text{qr}}} | \{\theta(l)\} \rangle\right\} \\
&=&\int \mathcal{D}\theta \left(\prod_{l=0}^{M-1} \prod_{i} e^{\frac{1}{\Delta \tau U}\cos(\theta_i(l)-\theta_i(l+1))} \right) \left(\prod_{l=0}^{M-1} e^{\Delta \tau t_{b} \sum_{\langle i, j\rangle} \cos (\theta_{i}(l)-\theta_{j}(l))}\right).
\end{eqnarray}

\section{Determinantal quantum Monte Carlo}
The determinantal quantum Monte Carlo(DQMC) is designed to deal with the interacting fermion lattice with quartic interactions and to decouple the quartic interactions into auxiliary bosonic fields coupled with fermion bilinears~\cite{BSS1981,Xu2017,WLJiang2021}. 

We perform DQMC  for the partition function
\begin{eqnarray}
\begin{split}
  Z=&\text{Tr}\{\prod_{m=1}^M e^{-\Delta\tau \hat H}\}
\end{split}
\label{eq:eqs7}
\end{eqnarray}
with $\hat H = \hat H_{\text{qr}} + \hat H_{f} + \hat H_{int}$. For small $\Delta\tau$, we make a similar approximation as in the previous section and take
\begin{eqnarray}
 Z&=&\text{Tr} \left\{\int \mathcal{D}\theta \prod_{l=0}^{M-1} \langle\{\theta(l+1)\} | e^{-\Delta \tau \hat H_{\text{qr}}} e^{-\Delta \tau \hat H_f} e^{-\Delta \tau \hat H_{int}} | \{\theta(l)\} \rangle\right\} \\
 &=&\int \mathcal{D}\theta \left(\prod_{l=0}^{M-1} \prod_{i} e^{\frac{1}{\Delta \tau U}\cos(\theta_i(l)-\theta_i(l+1))} \right) \left(\prod_{l=0}^{M-1} e^{\Delta \tau t_{b} \sum_{\langle i, j\rangle} \cos (\theta_{i}(l)-\theta_{j}(l))}\right) \text{Tr} \left\{\prod_{l=0}^{M-1} e^{-\Delta \tau \hat H_f} e^{-\Delta \tau \hat H_{\text{qr}-f}} \right\}.
\end{eqnarray}\label{eq:eqs8} 
We decompose the $Z$ into bosonic and fermionic part:
\begin{equation}
 Z= \int \mathcal{D}\theta \ \text{W}_b(\{ \theta \}) \text{W}_f(\{ \theta \})
\end{equation} \label{eq:eqs10}
where
\begin{eqnarray}
W_b &=& \left(\prod_{l=0}^{M-1} \prod_{i} e^{\frac{1}{\Delta \tau U}\cos(\theta_i(l)-\theta_i(l+1))}\right)\left(\prod_{l=0}^{M-1} e^{\Delta \tau t_{b} \sum_{\langle i, j\rangle} \cos (\theta_{i}(l)-\theta_{j}(l))}\right) \\
W_f &=& \text{Tr} \left\{\prod_{l=0}^{M-1} e^{-\Delta \tau \hat H_f} e^{-\Delta \tau \hat H_{int}} \right\}\\
&=&\det(\mathbf{1}+ \prod_{l=0}^{M-1} e^{-\Delta \tau H_f} e^{-\Delta \tau H_{int {\{ \theta(l) \}}}}) \\
&=&\det(\mathbf{1}+B(\beta,0)_{\{ \theta \}}).
\end{eqnarray}\label{eq:eqs9}
We sample the configuration of $\{ \theta \}$ and implement both local and Wolff cluster update schemes to avoid critical slowing down~\cite{WLJiang2019}.

In practice, we take the $\Delta\tau=0.1$ and sample about 2000 sweeps for each point.

\section{Absence of the sign problem}
The fermionic part of the Hamiltonian can be rewritten as $\hat{H}_{f}=-t_1\sum_{\left<i,j \right> \sigma,\lambda}\hat{c}_{i\sigma\lambda}^{\dagger}\hat{c}_{j\sigma\lambda}-t_2\sum_{\left<\left<i,j\right>\right>,\sigma,\lambda}\hat{c}_{i\sigma\lambda}^{\dagger}\hat{c}_{j\sigma\lambda} +h.c. = \hat{c} T_{ij}^{\sigma \lambda} c$ and $\hat{H}_{int} =-\frac{K}{2}\sum_{i \lambda}\hat{c}_{i \lambda}^{\dagger}\boldsymbol{\sigma}\hat{c}_{i \lambda}\cdot  \boldsymbol{\hat\theta}_i = \hat{c}V_{ii}^{\sigma \lambda}c$. We define the anti-unitary transformation
$\mathcal{K}=i \sigma_y K$ where $\sigma_y$ is Pauli matrix and $K$ is the complex conjugation operator. Under this anti-unitary transformation,
\begin{eqnarray}
\begin{split}
	\mathcal{K}T_{ij}^{\sigma \lambda}\mathcal{K}^{-1} &= T_{ij}^{\sigma \lambda}  \\
	\mathcal{K}V_{ii}^{\sigma \lambda}\mathcal{K}^{-1} &= -\frac{K}{2}\sum_{i}(\mathcal{K}\sigma_{x}\mathcal{K}^{-1}\cos(\theta) + \mathcal{K}\sigma_{y}\mathcal{K}^{-1}\sin(\theta)) \\	
	&=-\frac{K}{2}\sum_{i}(i \sigma_y \sigma_{x} (-i\sigma_y)\cos(\theta) + i\sigma_{y}\sigma_{y}(-i\sigma_{y})\cos(\theta)) \\
	&=	-\frac{K}{2}\sum_{i}(\sigma_y \sigma_{x} \sigma_y \cos(\theta)+ \sigma_{y}\sin(\theta)) \\
	&= -\frac{K}{2}\sum_{i}(\sigma_{x} \cos(\theta)+ \sigma_{y}\sin(\theta)) = V_{ii}^{\sigma \lambda}
\end{split}
\end{eqnarray}
Thus the fermion Hamiltonian is invariant under anti-unitary transformation $\mathcal{K}$.

In the presence of the antiunitary symmetry the fermion Hamiltonian is sign problem free. Within the DQMC framework, the fermion matrix is block diagonal in orbital index $\lambda$, so the Green function $G^{\lambda}(\tau,\tau) = (1+B^{\lambda}(\tau,0)B(\beta,\tau))^{-1}$ is also block diagonal in orbital index $\lambda$. The hamiltonian is an identity matrix in the orbital index and two block diagonal fermion matrix are identical.

\section{Controlling finite-size effects and the Critical point}
In practice, continuous field simulations of DQMC have very strong finite size effect. Therefore, it is necessary to reduce the finite size effect in simulations to save computing time and resource. 
To increase the momentum resolution on finite size simulations, we add up quantized magnitude to the system. The magnetic field is perpendicular to the lattice plane, therefore called $z$-direction flux. For free fermionic system, the magnetic field 
make the density of state smooth to gradually approach to the infinite system~\cite{Assaad2002}. The magnetic field is introduced via the Peierls phase factors on the bonds,
\begin{equation}
   \hat c_{i\sigma \lambda }^ + {{\hat c}_{j\sigma \lambda }} \rightarrow e^{i\int_{\mathbf{r}_i}^{\mathbf{r}_j} \mathbf{A}_{\sigma \lambda}(\mathbf{r}) d\mathbf{r}} \hat c_{i\sigma \lambda }^ + {{\hat c}_{j\sigma \lambda }} = e^{iA_{ij}} \hat c_{i\sigma \lambda }^ + {{\hat c}_{j\sigma \lambda }} 
\end{equation}   
with $\mathbf{B}=\triangledown \times \mathbf{A}$ and $\Phi_0$ the flux quanta. We take Landau gauge $\mathbf{A}(\mathbf{r})=-B(y,0,0)$, which is independent of spin and layer index. Therefore, the leftward and rightward hopping has opposite sign, and for longitudinal hopping, $A=0$. Note that, for the hopping term crossing the boundary, to satisfy the translation symmetry, $\mathbf{A_{ij}}$ is dependent with $L$. The phases $\mathbf{A_{ij}}$ for nearest-neighbor hopping read,
\begin{equation}
   A_{ij}= \left\{
      \begin{aligned}
         &+\frac{2\pi}{\phi_0} B \cdot i_y, \leftarrow \text{hopping}\\
         &-\frac{2\pi}{\phi_0} B \cdot i_y, \rightarrow \text{hopping}\\
         &0,\uparrow,\downarrow \text{hopping}\\
         &+\frac{2\pi}{\phi_0} B\cdot L \cdot i_x, \uparrow \text{hopping(boundary crossing)}\\
         &-\frac{2\pi}{\phi_0} B\cdot L \cdot i_x, \downarrow \text{hopping(boundary crossing)}\\
      \end{aligned}  
   \right.
\end{equation}
For the next-nearest-neighbor hopping, the hopping term along the diagonal lines is introduced. One unit cell is devided into four equvalent sub-regions. To guarantee the magnetic field strength in each areas, we thus write the add the phase to the next-nearest-neighbor hopping term as follows,
\begin{equation}
   A_{ij}= \left\{
      \begin{aligned}
         &+\frac{2\pi}{\phi_0} B \cdot i_y, \swarrow \text{hopping}\\
         &-\frac{2\pi}{\phi_0} B \cdot i_y, \nearrow \text{hopping}\\
         &+\frac{2\pi}{\phi_0} B \cdot i_y, \nwarrow \text{hopping}\\
         &-\frac{2\pi}{\phi_0} B \cdot i_y, \searrow \text{hopping}\\
         &+\frac{2\pi}{\phi_0} B \cdot (L i_x-i_y), \swarrow \text{hopping(boundary crossing)}\\
         &-\frac{2\pi}{\phi_0} B \cdot (L i_x-i_y), \nearrow \text{hopping(boundary crossing)}\\
         &+\frac{2\pi}{\phi_0} B \cdot (L i_x+i_y), \nwarrow \text{hopping(boundary crossing)}\\
         &-\frac{2\pi}{\phi_0} B \cdot (L i_x+i_y), \searrow \text{hopping(boundary crossing)}\\
      \end{aligned}  
   \right.
\end{equation}
where $B=\frac{\Phi_0}{L^2}$ is the unit magnetic flux, and $i_x, i_y$ are the indices of site range between $1$ and $L$ in the $x$ and $y$ lattice directions. Various arrows represent the direction of hopping terms from site $i$ to $j$. Note that when $L \rightarrow \infty$, the magnetic field approaches 0, and the Hamiltonian goes back to the original one. However, the magnetic field breaks the translation symmetry, i.e., the momentum $k$ is not a good quantum number for fermion. In practice, the magnetic flux is only added when measuring bosonic observables, e.g., bosonic susceptibility, while for fermionic observables in which $k$-space resolution is needed, e.g. fermi surface and fermion self energy, we do not include the magnetic flux.
\begin{figure}[htp!]
\includegraphics[width=\textwidth]{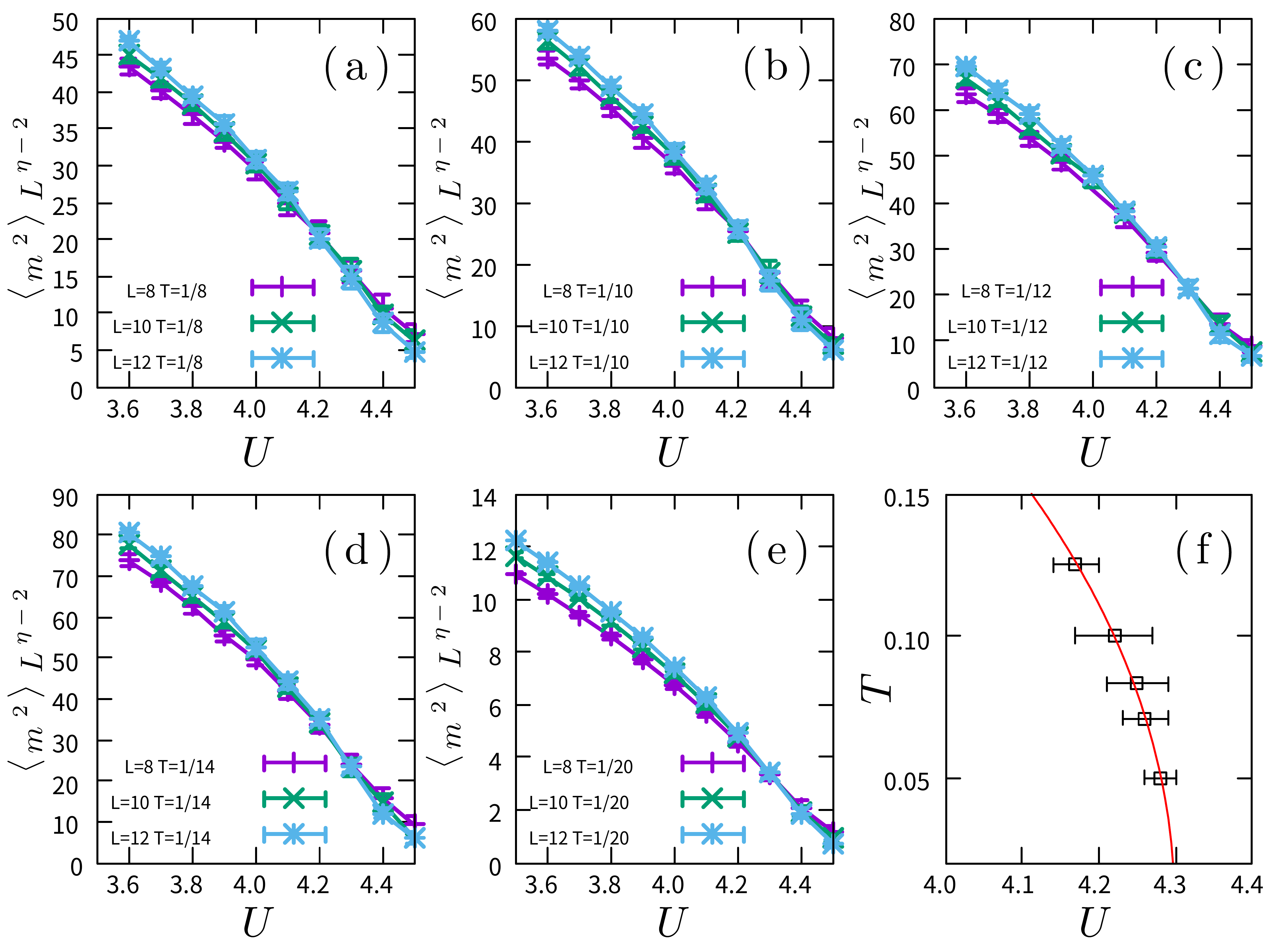}
\caption{(a) (b) (c) (d) (e) are bosonic order parameter under the different $T$ where cross point of different $L$ susceptibility is critical point. (f) is the phase diagram and we extrapolate critical point from finite temperature to zero temperature which is located in $U=4.30(3)$    }
\label{fig:figs2}
\end{figure}

We obtain the critical point by the bosonic parameter $\left\langle m^2 \right\rangle=\frac{1}{N}\sum_{i,j}\left\langle S_i S_j \right\rangle$.As is shown in Fig.\ref{fig:figs2}, We observe the bosonic parameter versus $U$ under the variant finite temperature, the $\eta = 0.25$ is corresponding to 2D XY phase transition. The cross point between different $L$ is identified as transition point.Fig.\ref{fig:figs2} is phase diagram which the boundary is obtained by bosonic order parameter and zero temperatrue critical point $U=4.30(3)$ can be extrapolated by the finite temperatrue data point.

\section{Fitting of QMC data}
We present \addYW{in} detail the comparison of QMC data with analytical theory. First of all, for a ferromagnetic boson-fermion model near the QCP, we predict the following fermionic and bosonic form~\cite{Chubukov2014}:
\begin{eqnarray}
D^{-1}(\textbf{q},\Omega_n) = D_0^{-1} (M^2 + |\textbf{q}|^2 + \kappa |\Omega_n|)
\end{eqnarray} \label{eq:eqs11}

The bosonic propagator is per real-space component $D_{ij} = \frac{1}{N_b}\delta_{ij}\left\langle S_i S_j \right\rangle$ with $N_b=2$ being the number of bosons.The susceptibility is defined the $\chi_{ij}=\delta_{ij}\left\langle S_i S_j \right\rangle$. So we have the ralation:

\begin{equation}
\chi^{-1} = \frac{1}{2} D^{-1}
\end{equation} \label{eq:eqs12}

\begin{figure}[htp!]
\includegraphics[width=\textwidth]{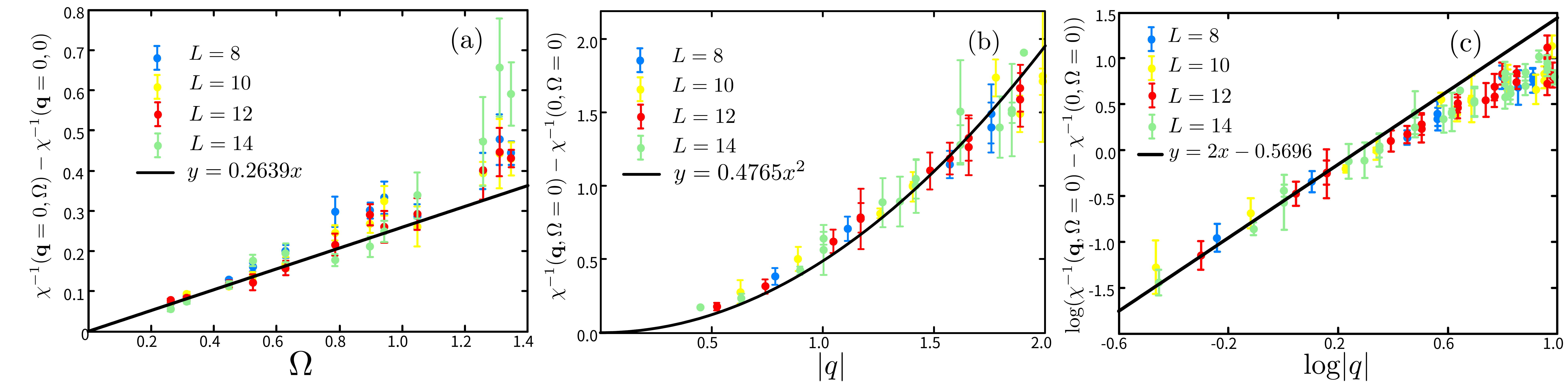}
\caption{QMC data fit for various $L$ at the QCP. (a) the inverse of bosonic susceptibility versus $\Omega_n$ when $\mathbf{q}=0$. (b) the inverse of bosonic susceptibility versus $\mathbf{q}$ when $\omega_n=0$ (c) the log-log relation for the inverse of bosonic susceptibility versus $\textbf{q}$, $\chi(|\mathbf{q}|,\omega_n =0) \sim |\mathbf{q}|^2$ relation is clearly seen.}
\label{fig:figs1}
\end{figure}

From the relation of Eq. S17, The parameter of $D_0$ and $\kappa$ can be fitted by QMC data as Fig.\ref{fig:figs1}. The $D^{-1}_0 \kappa$ is obtained by Fig.\ref{fig:figs1}(a) and $D_0 ^{-1}$ can be observed by Fig.\ref{fig:figs1}(b) which have the following results
\begin{eqnarray}
D^{-1}_0\kappa &=& 0.2639\times 2 = 0.5278 \\
D^{-1}_0 &=& 0.4765 \times 2 = 0.9530 \\
 \kappa &=& 0.5538
\end{eqnarray} \label{eq:eqs13} 

We derive the self energy $\Sigma(\textbf{k},\omega)$ from the fermionic Green function and modified Eliashberg theory(MET)~\cite{XYXu2020, AKlein2020}. The fermion Green function $G(\textbf{k},\omega)$ is given by
\begin{equation}
G(\textbf{k},\omega)=(i\omega_n + i\Sigma(\textbf{k},\omega)-\epsilon(\textbf{k}))^{-1}
\end{equation} \label{eq:eqs14}

At the QCP, we focus on the momentum point $\textbf{k}_F$ on the Fermi surface. The energy band appears the linear behavior near the $\textbf{k}_F$.So we have 
\begin{equation}
G^{-1}(\textbf{k},\omega_n)=i\omega_n + i\Sigma(\textbf{k},\omega_n)-\textbf{v}_F \cdot (\textbf{k}-\textbf{k}_F)
\end{equation} \label{eq:eqs15}
where $\textbf{v}_F$ is Fermi velocity at Fermi surface. The self energy can be calculated by the MET as the following form~\cite{XYXu2020,AKlein2020} 
\begin{eqnarray}
	-i\Sigma(\textbf{k})=\bar{g}N_{b}T\sum_{n}\int\frac{d^2 p}{(2\pi)^2}G(\textbf{p+k})D(\textbf{p})
\end{eqnarray} \label{eq:eqs16}
which $\bar{g}$ is the effective coupling and $T$ is the temperature.The bare fermion-boson vertex coupling is $\lambda=K/2=0.5$ in our model which the effective coupling is given by form 
\begin{eqnarray}
	\bar{g} = \lambda^{2}D_0 = 0.2623
\end{eqnarray} \label{eq:eqs17}

Combining the QMC data for self energy, we can assume the $|\Sigma| \ll \omega_n$. So the bare fermionic green function   $G(\textbf{k},\omega_n)=(i\omega_n -\textbf{v}_F \dot (\textbf{k}-\textbf{k}_F))^{-1}$ can be used in the Eq.S22 which the self energy $\Sigma(\textbf{k})$ appears

	\begin{equation}
	\Sigma(\textbf{k}_F,\omega_n) \approx N_b\bar{g}T\sum_{l}\int_{0}^{\infty}\frac{pdp}{2\pi}\frac{\sigma(\omega_l)}{\sqrt{\omega_l^2+
	\textbf{v}^{2}_F(\theta_k)p^2}}\frac{1}{M^2+p^2+\kappa|\omega_n-\omega_l|}
	\end{equation}\label{eq:eqs18}
where $\sigma(\omega_l)$ is the sign function, Plugging $\omega = \textbf{v}_F(\theta_k)p$ into Eq.S24 yields

	\begin{equation}
	\Sigma(\textbf{k}_F,\omega_n) \approx N_b\bar{g}T\sum_{l}\int_{0}^{\infty}\frac{\omega d\omega}{2\pi}\frac{\sigma(\omega_l)}{\sqrt{\omega_l^2+
	\omega^2}}\frac{1}{M^2\textbf{v}^2_F+\omega^2+\kappa\textbf{v}^2_F|\omega_n-\omega_l|}
	\end{equation}\label{eq:eqs19}

we set $\omega_c =\kappa\textbf{v}^2_F$ and the self energy  $\Sigma(\textbf{k}_F,\omega_n)$ is transformed into
	\begin{equation}
	\Sigma(\textbf{k}_F,\omega_n) \approx N_b\bar{g}T\sum_{l}\int_{0}^{\infty}\frac{\omega d\omega}{2\pi}\frac{\sigma(\omega_l)}{\sqrt{\omega_l^2+
	\omega^2}}\frac{1}{M^2\textbf{v}^2_F+\omega^2+\omega_c|\omega_n-\omega_l|}
	\end{equation}\label{eq:eqs20}
	
	According to the conclusion of our previous works~\cite{XYXu2020,AKlein2020}, $|\Sigma(\omega_n)| \ll \omega_n$ will lead to the $\Sigma(\omega_n)$ splitting for thermal and quantum part. The quantum part is the zero-temperature fermionic self-energy and the thermal part carries a very simple form  $1/\omega_n$. At small $\omega_n$, the large thermal contribution makes it hard to detect the quantum part.sWe separate the quantum part by cutting off the thermal part which the self energy can be transformed into
	\begin{equation}
	\Sigma(\omega_n) = \Sigma_T(\omega_n,T \neq 0) + \Sigma_Q(\omega_n,T)
	\end{equation}\label{eq:eqs21}
	where $\Sigma_T$ is the $\omega_l=\omega_n$ piece of the sum in Eq.S26. namely
	
	\begin{equation}
	\Sigma_T(\omega_n)\approx\frac{\bar{g}T}{2\pi\omega_n}S(\frac{\upsilon_F M}{|\omega_n|})
	\end{equation}\label{eq:eqs22}
	where 
	\begin{equation}
	S(x)=\left\{
	\begin{array}{rcl}
	\frac{\text{acosh}(1/x)}{\sqrt{1-x^2}} & & {x<1} \\
	\frac{\text{acos}(1/x)}{\sqrt{x^2-1}} & & {x>1} 
	\end{array}
	\right.
	\end{equation}\label{eq:eqs23}
	As $S(x)$ vanishes rapidly at large $x$, it predicts that $\Sigma_T$ only contributes significantly at finite temperatrue and close enough to the QCP. $\alpha(T,\omega)=\omega_n\Sigma_T(\omega_n)$ depends on frequency at the small $\omega_n$. $\alpha(T,\omega_n) \approx \alpha(T)$
	
The quantum part includes all the terms in the Matsubara sum when $T \rightarrow 0$ The sum can be replaced by the integral which the form of self energy is 

    \begin{equation}
	\Sigma_Q=\bar{g}\sigma(\omega_n)(\frac{\omega_n}{•\omega_c})^{1/2}u(\frac{\omega_n}{\omega_c})
	\end{equation}\label{eq:eqs24}
	with
	\begin{equation}
	u(z) = \int_{0}^{\infty}\frac{dxdy}{4\pi^2}\frac{1}{x^2+y}(\frac{\sigma(y+1)}{\sqrt{1+(\frac{y+1}{x})^2 z^2}}-\frac{\sigma(y-1)}{\sqrt{1+(\frac{y-1}{x})^2 z^2}})
	\end{equation}\label{eq:eqs25}
Because the $u(z)\rightarrow \frac{1}{2\pi}$ when $z \rightarrow 0$, the quantum part have $\omega_n^{1/2}$ asymptotic behavior at the smallest $\omega_n$

\end{document}